\documentclass[ twocolumn, twocolappendix]{aastex631}

\definecolor{purple1}{rgb}{0.6, 0.0, 0.8}
\definecolor{green1}{rgb}{0.25, 0.5, 0.25}
\definecolor{red1}{rgb}{0.7, 0.15, 0.15}
\definecolor{blue1}{rgb}{0.11, 0.09, 0.71}

\shorttitle{Emission from Shock-Ionized Jets}
\shortauthors{Gardiner et al.}

\graphicspath{{./}{old_figs/}}

\begin{document}

\title{Disk Wind Feedback from High-mass Protostars. IV. Shock-Ionized Jets}

\author[0000-0002-8857-613X]{Emiko C. Gardiner}
\affiliation{Department of Astronomy, University of California at Berkeley, 94720, Berkeley, CA, USA}

\author[0000-0002-3389-9142]{Jonathan C. Tan}
\affiliation{Department of Space, Earth and Environment, Chalmers University of Technology, SE-41296 Gothenburg, Sweden}
\affiliation{Department of Astronomy, University of Virginia, Charlottesville, VA 22904, USA}

\author{Jan E. Staff}
\affiliation{Department of Space, Earth and Environment, Chalmers University of Technology, SE-41296 Gothenburg, Sweden}
\affiliation{College of Science and Math, University of the Virgin Islands, St Thomas, 00802, United States Virgin Islands}

\author[0000-0002-3835-3990]{Jon P. Ramsey}
\affiliation{Department of Astronomy, University of Virginia, Charlottesville, VA 22904, USA}

\author{Yichen Zhang}
\affiliation{Department of Astronomy, University of Virginia, Charlottesville, VA 22904, USA}

\author{Kei E. I. Tanaka}
\affiliation{Department of Earth and Planetary Sciences, Tokyo Institute of Technology, Meguro, Tokyo, 152-8551, Japan}



\begin{abstract}
Massive protostars launch accretion-powered, magnetically-collimated outflows, which play crucial roles in the dynamics and diagnostics of the star formation process. Here we calculate the shock heating and resulting free-free radio emission in numerical models of outflows of massive star formation within the framework of the Turbulent Core Accretion model. We post-process 3D magneto-hydrodynamic simulation snapshots of a protostellar disk wind interacts with an infalling core envelope, and calculate shock temperatures, ionization fractions, and radio free-free emission. We find heating up to $\sim10^7\:$K and near complete ionization in shocks at the interface between the outflow cavity and infalling envelope. However, line-of-sight averaged ionization fractions peak around $\sim$10\%, in agreement with values reported from observations of massive protostar G35.20-0.74N. By calculating radio continuum fluxes and spectra, we compare our models with observed samples of massive protostars. 
We find our fiducial models produce radio luminosities similar to those seen from low and intermediate-mass protostars that are thought to be powered by shock ionization. Comparing to more massive protostars, we find our model radio luminosities are $\sim10$ to 100 times less luminous. We discuss how this apparent discrepancy either reflects aspects of our modeling related to the treatment of cooling of the post-shock gas or a dominant contribution in the observed systems from photoionization. Finally, our models exhibit 10-year radio flux variability of $\sim$5\%, especially in the inner 1000 au region, comparable to observed levels in some hyper-compact HII regions.
\end{abstract}



\section{Introduction}
\label{sec:intro}

Massive stars, i.e., those with mass $\gtrsim 8\:M_\odot$, have a profound influence on the evolution of the universe via their radiative, mechanical and chemical feedback. However, the formation mechanism of massive stars remains under active debate, with two main theoretical paradigms being considered: \textit{Core Accretion} and \textit{Competitive Accretion}. Core Accretion is a scaled-up version of the standard model for low-mass star formation, in which self-gravity drives the formation of a concentrated core from which matter then accretes to a central protostar \citep{Shuetal1987}. The Turbulent Core Accretion (TCA) model \citep{McKeeTan2003} extends this model by setting the boundary conditions of the initial pre-stellar core to pressure and virial equilibrium with a surrounding protocluster clump environment. Alternatively, Competitive Accretion \citep[e.g.,][]{2001MNRAS.323..785B,Wangetal2010,2022MNRAS.512..216G} proposes that massive stars form as part of a protocluster that globally funnels material to its central regions. Here protostars compete for the gas, with accretion proceeding in a chaotic manner without the presence of a massive, coherent core. Numerical simulations of competitive accretion generally predict that massive stars form relatively slowly compared to the TCA model, since the mass needs to be accumulated from larger scales.

While massive star formation models remain under debate, observations of their outflows can help constrain these models. One can simulate the outflow from a massive protostar under a given accretion scenario, and compare its predicted properties to observed systems. Understanding the launching and propagation of these outflows is crucial not only for testing formation models, but also to understanding how massive protostars impact their surrounding core and clump environments \citep[e.g.,][]{Arceetal2007}. 




In this paper series we simulate disk-wind \citep{BlandfordPayne1982} feedback from a massive protostar forming from a $60\:M_\odot$ core in the context of the Turbulent Core Accretion (TCA) model of massive star formation \citep{McKeeTan2003}. In Paper I \citep{Staffetal2019}, MHD simulations at a given evolutionary stage were presented and basic properties, such as outflow cavity geometry and star formation efficiency were investigated. In Paper II \citep{Staffetal2023}, a full evolutionary sequence via a single simulation that followed growth of the protostar from low to high masses was presented. In Paper III (Xu et al. 2023, sub.) predictions for CO line emission from these simulations have been calculated. 

Here, in Paper IV we present a model for the expected ionization due to shocks in the simulation and predictions for the associated radio continuum emission.
The structure of the paper is as follows. In \S\ref{sec:methods} we describe details of the simulations, calculations of shock parameters, and methods for predicting emission. In \S\ref{sec:results} we present the simulation shock-modeling results and compare them to observations. In \S\ref{sec:discussion} we discuss the caveats and limitations of the modeling and present the main conclusions of our study.

\section{Methods}
\label{sec:methods}
\subsection{MHD Simulations}
\label{sec:meth_simulation}

The post-processing analysis that we present in this paper uses snapshots from a 3D, ideal magneto-hydrodynamic (MHD) simulation of a protostellar outflow interacting with its surrounding natal core infall envelope. The simulation domain includes one hemisphere of the protostellar core, from $100\:{\rm au}$ above the accretion disk mid-plane, up to a height of $25,000\:{\rm au}$ \citep[see][for details]{Staffetal2023}. The outflow is injected into the simulation box at the lower $z$ boundary. Mass can accrete from the envelope by flowing out of the lower $z$ boundary, and this accreted mass results in the star and accompanying accretion disk growing \citep{Staffetal2023}. The rate at which mass flows through the $z$ boundary is constrained so that the growth rate of the star matches the expectations of the TCA model, i.e., specifically the results of \citet{Zhangetal2014_paperIII}. Outflow boundary conditions allowing mass to flow out are used at all the other boundaries.

As the star grows, the mass and momentum input flux rates of the outflow also change with time, again following \citet{Zhangetal2014_paperIII}. Initially, the envelope has a mass of $60\:{M_\odot}$ and a radius of $\sim12,000\:{\rm au}$. The envelope is initialized with a power-law density profile of the form $\rho \propto r^{-3/2}$ \citep{McKeeTan2003}, and is unstable to gravitational collapse. The simulation starts with a $1\:{M_\odot}$ protostar at the center of the envelope. The envelope begins to collapse under the force of gravity, and its accretion and the propagation of the outflow are simulated for about 100,000 years, by which point the star has grown to just over $25\:{M_\odot}$.

The simulation was run using ZEUS-MP \citep{Normanetal2000}, with an isothermal equation of state and a fixed sound speed of $0.9\:{\rm km~s^{-1}}$. The simulation used a logarithmically stretched grid in all three dimensions, consisting of $168 \times 280 \times 280$ cells. The cells are smallest near the outflow axis and near the lower $z$ boundary, where the minimum cell size is $\sim$12 au.
The initial core is threaded by a ``Blandford-Payne'' like poloidal magnetic field \citep{BlandfordPayne1982} plus a constant vertical component added to it to ensure a core flux of $\sim 1\:{\rm mG}$.


\subsection{Calculating Shock Temperatures, Ionization Fractions and Free-Free Radio Emission} 
\label{sec:meth_shocks}

A general overview of our method is the following. In order to model shock-ionization and the resulting emissions, we first calculate the shock velocities between cells and then use these to calculate post-shock temperatures (which is required because the MHD simulation is isothermal). Then, these temperatures are used to calculate the ionization fractions expected in the post-shock gas. Two cases for the volume filling factor of the post-shock gas have been considered. Case A assumes the post-shock gas fills the entire cell into which the gas is flowing (averaging over all incoming flows, i.e., up to six). Case B, which is our fiducial model, estimates a cooling time in the post-shock gas and then assumes the filled volume is only a fraction of the cell equal to the ratio of cooling time to flow crossing time.
Then, given the temperatures and ionization fractions, we evaluate the free-free emissivity in the radio regime and calculate the resulting radio spectrum from the computational domain.
The details of these methods are explained in Appendix~\ref{app:method}.




\section{Results}
\label{sec:results}

\subsection{Shock Structures}
\label{sec:results_shocks}

\begin{figure*}
    \centering
        \includegraphics[width=0.49\linewidth]{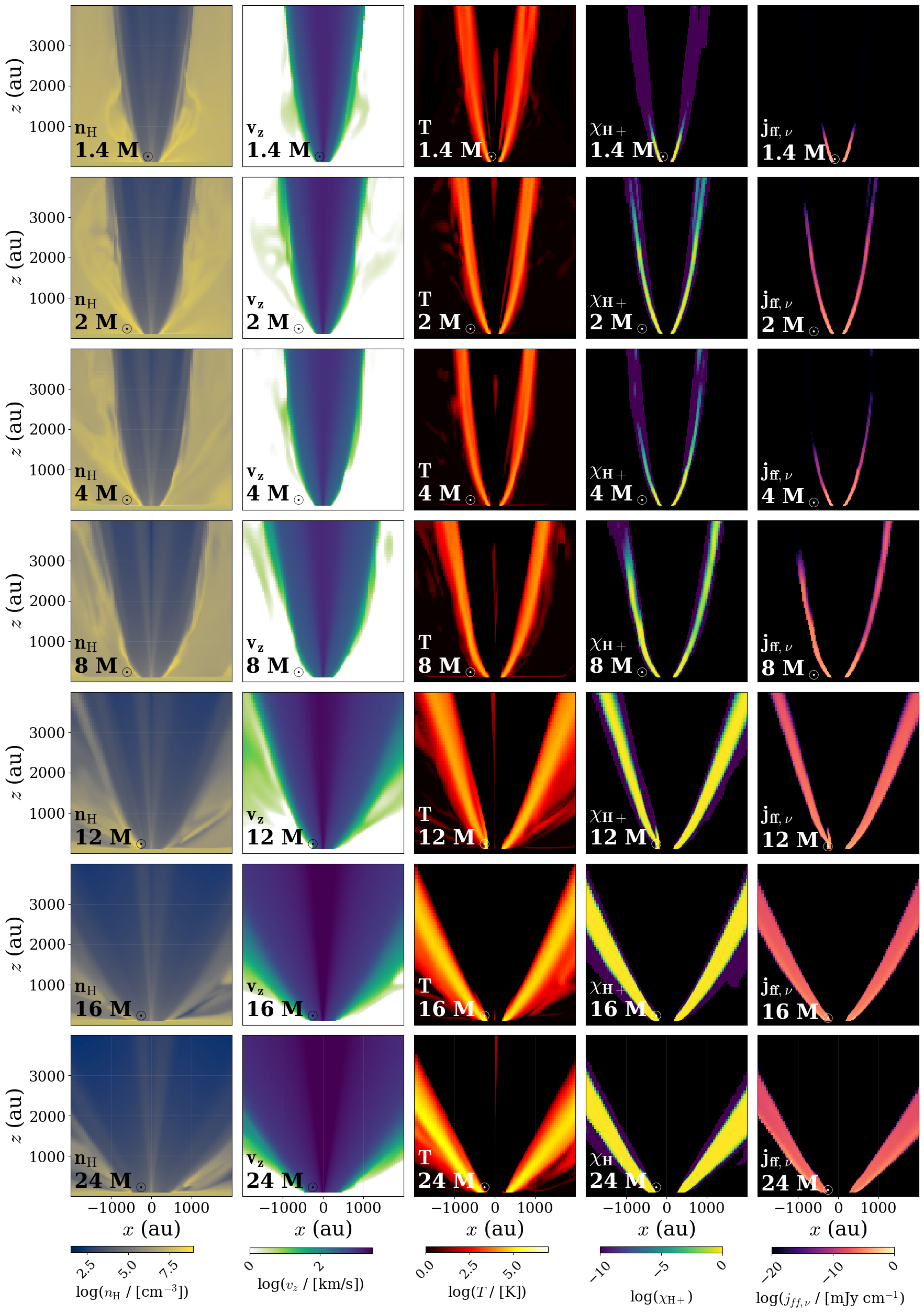}
        \includegraphics[width=0.49\linewidth]{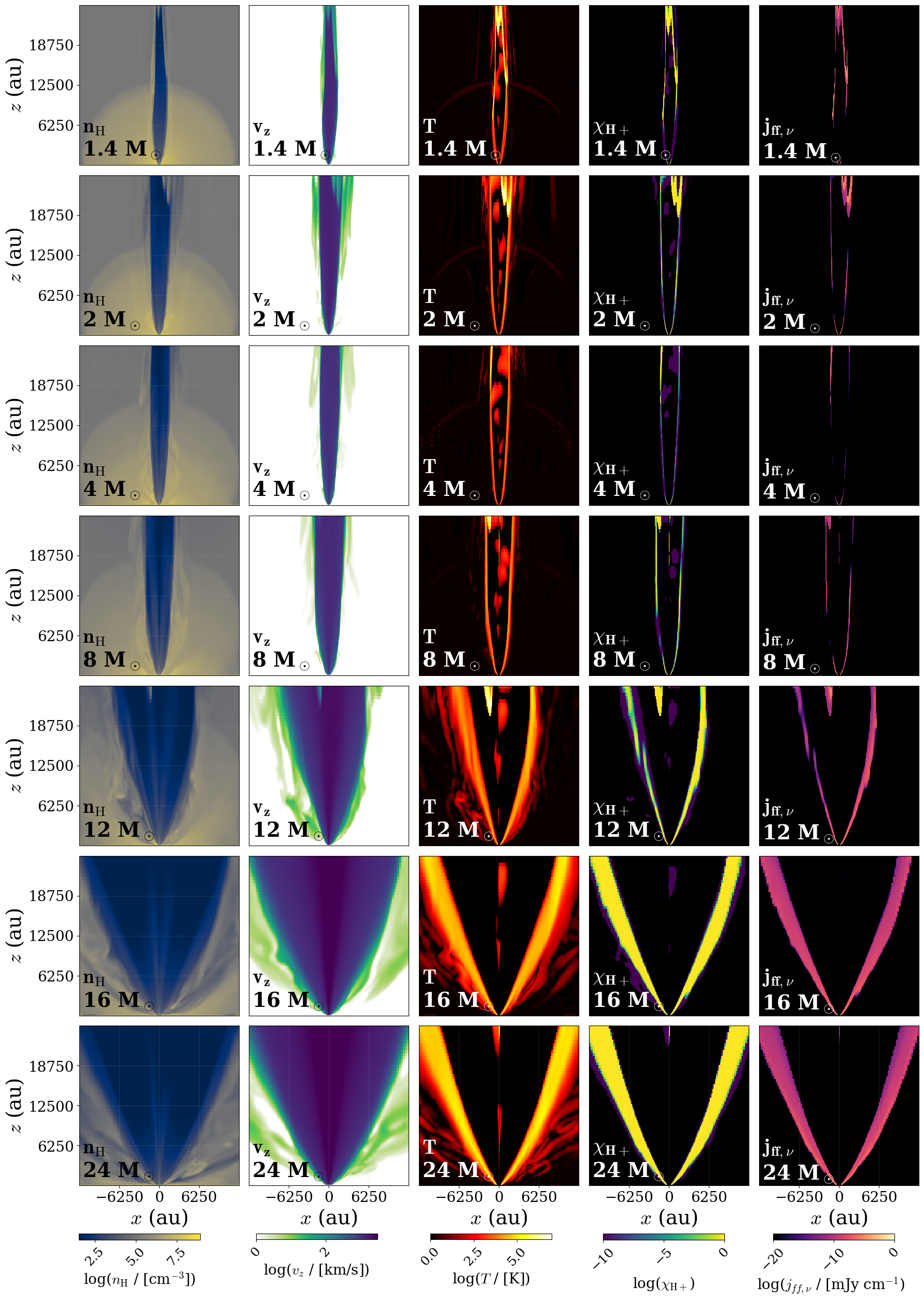}
        \caption{Time evolution of shock structures in slices through the center of the simulation in the $x-z$ plane, over the ranges (\textit{left panel}) $-2000\:\mathrm{au} < x < 2000\:\mathrm{au}$, $0\:\mathrm{au} < z < 4000\:\mathrm{au}$ and (\textit{right panel}) $-12500\:\mathrm{au} < x < 12500\:\mathrm{au}$, $0\:\mathrm{au} < z < 25000\:\mathrm{au}$. From left to right, the columns display number density, velocity in the z-direction, shock-heated temperature, ionization fraction, and free-free emissivity. Rows from top to bottom show the evolution of the simulation with protostellar masses of 1.4, 2, 4, 8, 12, 16, and 24~$M_\odot$ at 4,000, 9,000, 21,000, 39,000, 54,000, 68,000, and 94,000 years, respectively.\label{fig:slices}}
\end{figure*}

Snapshots from the simulation at various protostellar masses, ranging from 1.4 to 24~$M_\odot$ (corresponding to times from 4,000 to 94,000 yr)
were selected for post-processing to calculate shock heated temperatures, ionization fractions and radio free-free emission.
Examples of the shock structures in a slice through the center of these snapshots ($y = 0$) are displayed in Fig.\ \ref{fig:slices} for both an inner 4,000~au scale and the global 25,000~au scale. The density and z-velocity magnitude produced from the simulation are displayed in the first and second columns, respectively. The density map reveals a central low-density cavity, i.e., with $n_{\rm H}\sim 10^{0}-10^{2}\:$cm$^{-3}$, while the surrounding regions in the infall envelope reach densities of $n_{\rm H}\sim 10^{7}-10^{9}\:$cm$^{-3}$. This cavity increases in opening angle as it evolves, especially in the later stages from 8 to 24~$M_\odot$. The $z$-velocities exceed 1000 km/s in the outflow cavity, but become much lower in the surrounding regions where the outflow is interacting with the infall envelope.

The third column shows the mass flux weighted post-shock temperatures. Temperatures are highest, up to $\sim 10^7$ K, on the boundary between the low-density, high-velocity outflow and the surrounding higher-density, low-velocity envelope. The fourth column shows these high-temperature regions are nearly fully ionized. Finally in the fifth column the emissivities, $j_{\nu,\mathrm{ff}}$ at 5.3~GHz are shown.

\subsection{Ionization Fractions}
\label{sec:results_ionfrac}

\begin{figure*}
    \centering
    \includegraphics[width=1.0\linewidth]{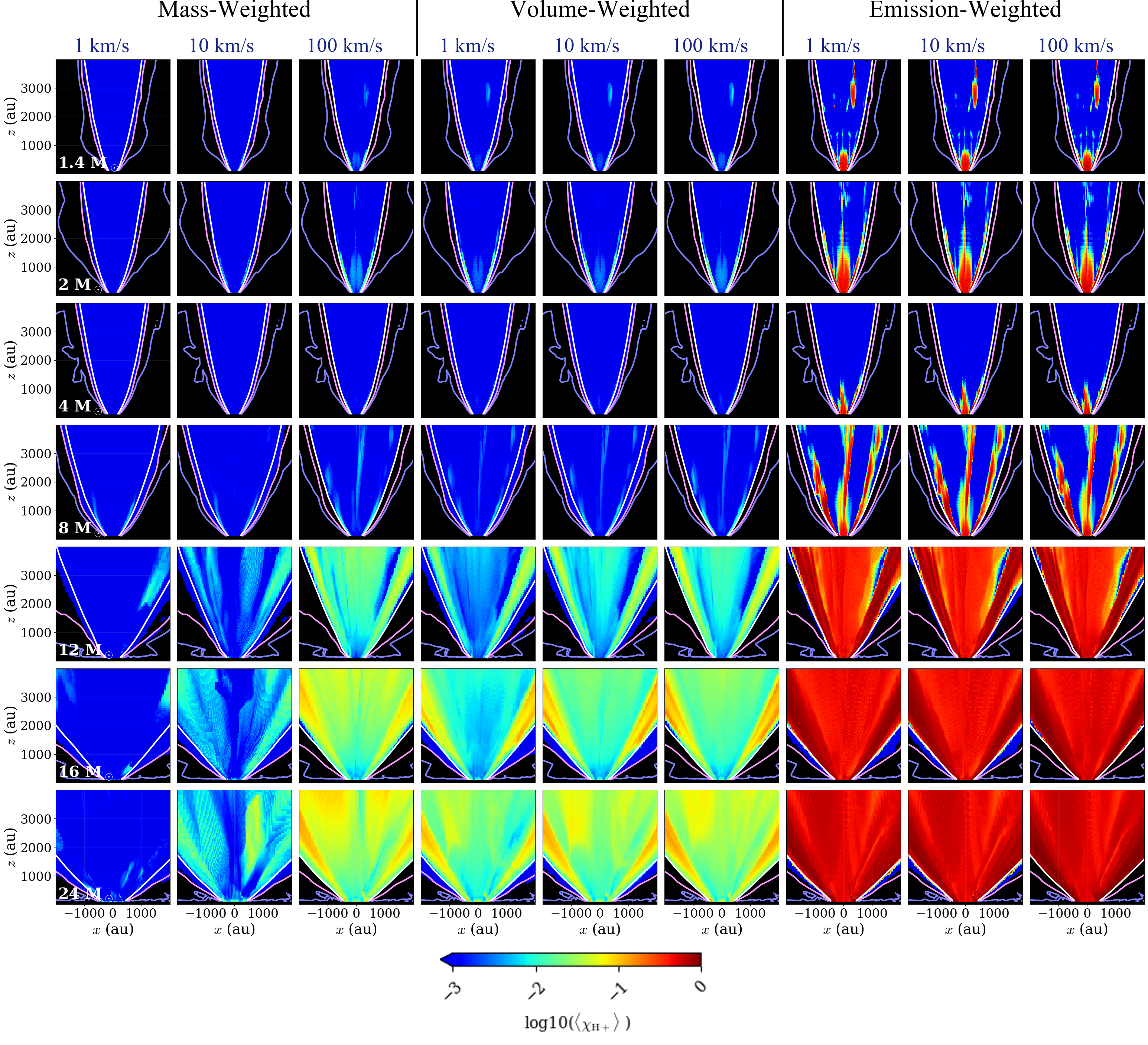}
    \caption{Average ionization fractions weighted by mass (\textit{first three columns}), volume (\textit{middle three columns}), and emission (\textit{last three columns}) for Case B (with cooling). The averages are calculated over all cells in the y-column that exceed cutoff velocities of $v_z \geq$ 1 km/s (\textit{columns 1, 4, 7}), 10 km/s (\textit{columns 2, 5, 8}), and 100 km/s (\textit{columns 3, 6, 9}), plotted over  $-2000\:\mathrm{au} < x < 2000\:\mathrm{au}$, $0\:\mathrm{au} < z < 4000\:\mathrm{au}$. From top to bottom, the simulation grows in mass, reaching 1.4, 2, 4, 8, 12, 16, and 24~$M_\odot$ at 4,000, 9,000, 21,000, 39,000, 54,000, 68,000, and 94,000 yr, respectively. Contours are overlaid displaying where the maximum z-velocity along the projection meets the cutoffs of 1 km/s (\textit{white}), 10 km/s (\textit{pink}), and 100 km/s (\textit{purple}). 
    \label{fig:chimaps_4000_ratio}}
\end{figure*}

\begin{figure*}
    \centering
    \includegraphics[width=1.0\linewidth]{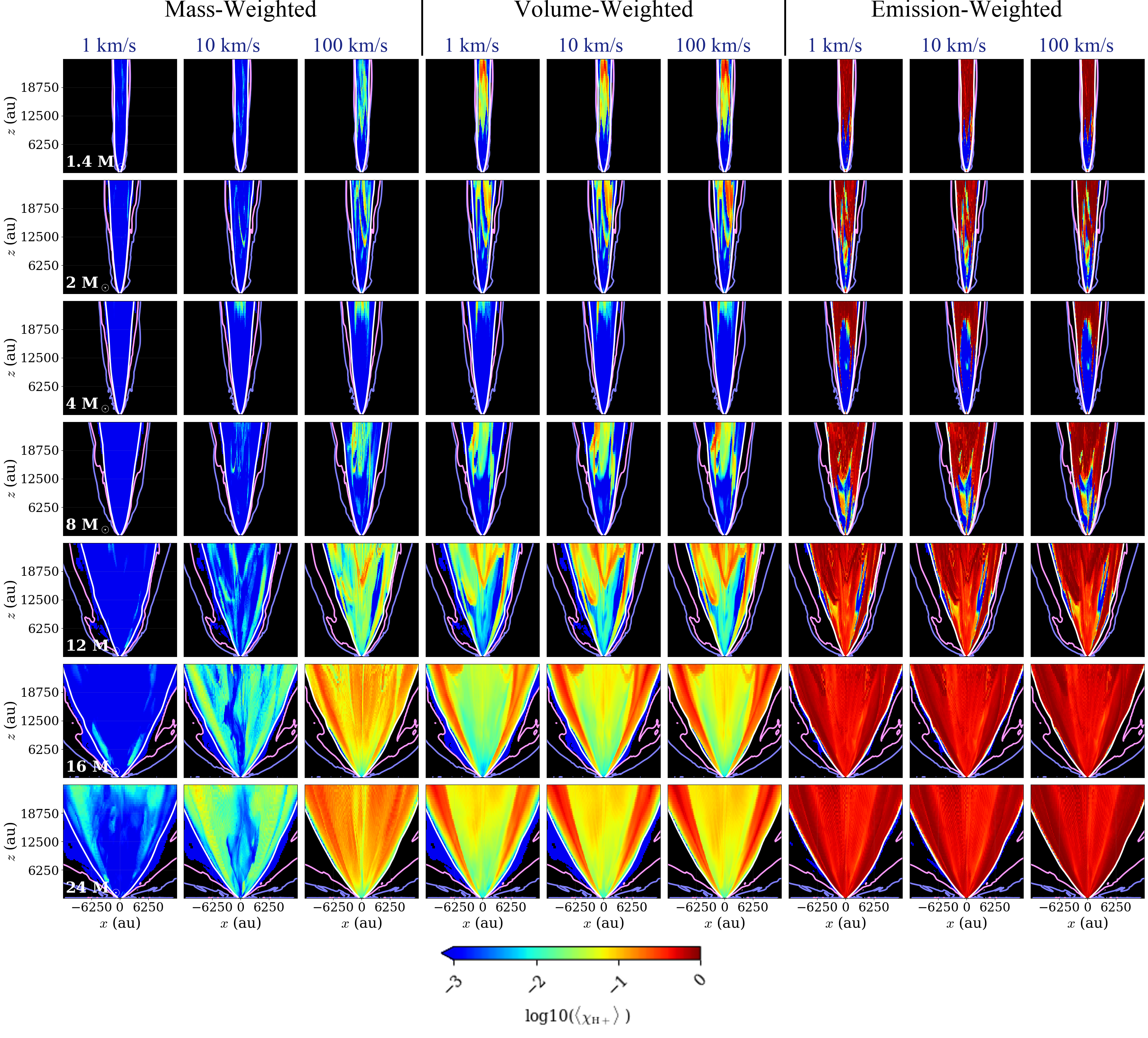}
    \caption{Same as Fig.\ \ref{fig:chimaps_4000_ratio}, but for $-12500\:\mathrm{au} < x < 12500\:\mathrm{au}$, $0\:\mathrm{au} < z < 25000\:\mathrm{au}$.\label{fig:chimaps_25000_ratio}}
\end{figure*}

Ionization fraction in the outflow is one metric by which the results of our model can be compared to observations. The outflowing material was defined as including any cell with $v_z > v_\mathrm{min}$, for $v_\mathrm{min}$ = 1, 10, and 100 km s$^{-1}$. 
We averaged the ionization fractions over the line-of-sight ($y$) direction considering three different ways of doing the averaging: (1) mass-weighted, (2) volume-weighted; (3) emission-weighted.
The resulting ionization fraction maps are shown in Fig. \ref{fig:chimaps_4000_ratio} and \ref{fig:chimaps_25000_ratio} for the 4,000~au and 25,000~au scales, respectively. The overall ionization fractions averaged by column density, area, and intensity weighting over the mass, volume, and emission-weighted maps, respectively, are summarized in Table~\ref{tab:avg_chi_ratio}. See Figs. \ref{fig:chimaps_4000_noratio} and \ref{fig:chimaps_25000_noratio} and Table~\ref{tab:avg_chi_noratio} for the equivalent Case A results.


Higher velocity cutoffs lead to higher average ionization fractions because they exclude slower-moving, less ionized gas, particularly on the outskirts of the outflow. Weighting by radio free-free emissivity provides the highest and most high-contrast ionization fractions because, with ionization fraction and emissivity both dependent on shock temperatures and emissivity depending on $\chi_\mathrm{H+}^2$, high emissivity correlates to high ionization fraction and vice versa. 



For our fiducial case, Case B, which considers cooling, only the ionization from the portion of each cell flooded by the shock before cooling takes place is accounted for, leading to a reduction in the estimated ionization fractions, except for the emission-weighted metric.

\begin{figure*}
\includegraphics[width=1.0\textwidth]{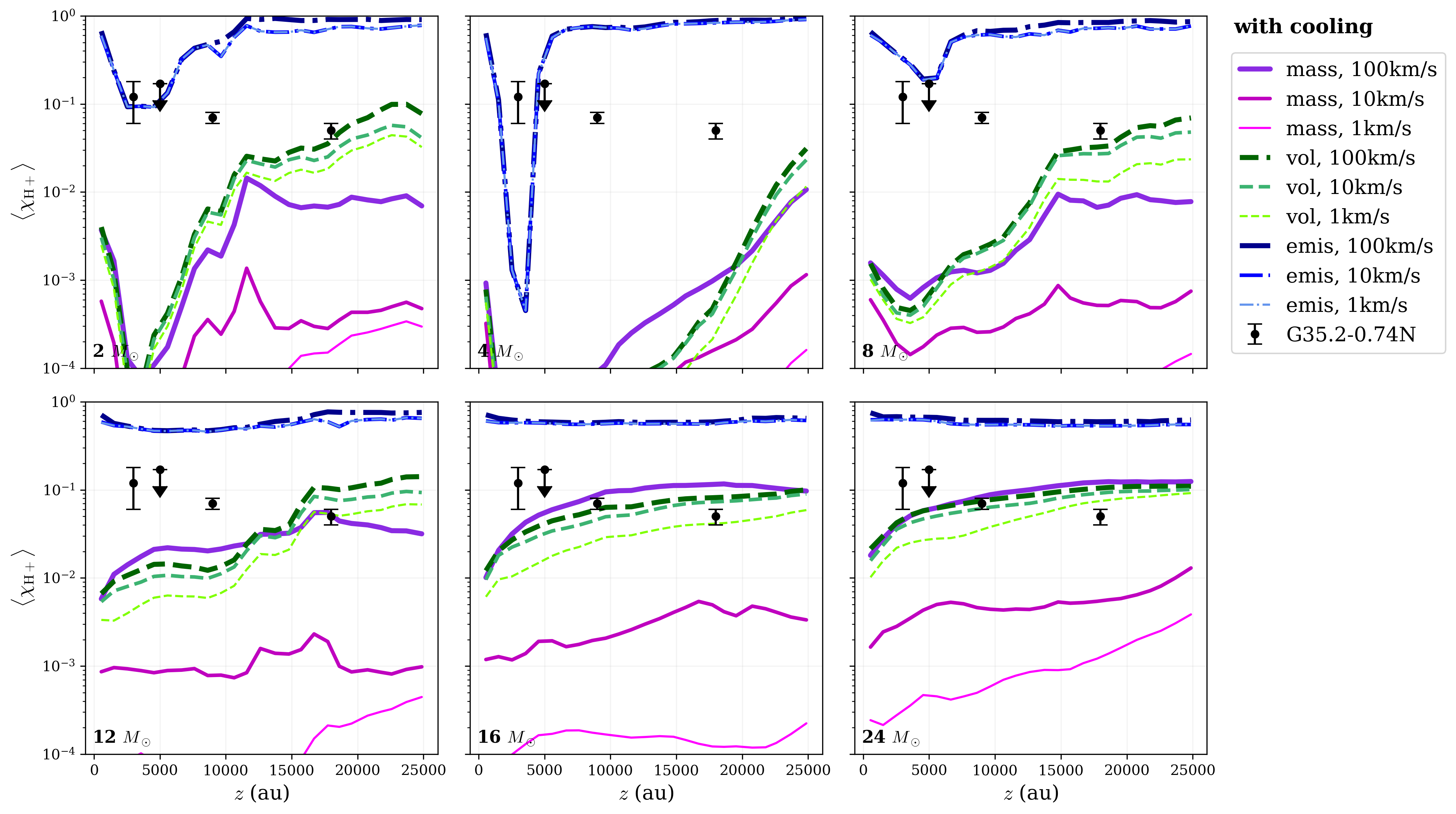}
\caption{Average ionization fractions versus height for Case B (with cooling) for $m_*=$ 2, 4, 8, 12, 16, and 24~$M_\odot$ snapshots, as labeled.
The ionization fractions were averaged by mass (\textit{solid pink}), volume (\textit{dashed green}), and emission (\textit{dash-dotted blue}) over 1000 au wide regions along the z-direction and for different cutoff velocities of $v_z \geq$ 100 km/s (\textit{thick}), 10 km/s (\textit{medium}), and 1 km/s (\textit{thin}). Overlaid are the ionization fractions calculated in 4 observed knots in the outflow from massive protostar G35.2-0.74N \citep{Fedrianietal2019}.
\label{fig:chiprofiles_ratio}}
\end{figure*}

Figure~\ref{fig:chiprofiles_ratio}
shows the Case B profiles of ionization fraction along the outflow axis, averaging over strips of width $\Delta z$ = 1000~au (see Fig. \ref{fig:chiprofiles_noratio} for Case A). Here we more clearly see that emission-weighting generally results in the highest ionization fractions, where regions above $\sim$4000 au are almost entirely ionized. Dips are present near 2000-4000 au, however, corresponding to a lack of shock-emission (see the emissivity column of Fig.\ \ref{fig:slices}). Higher velocity cutoffs correspond to higher ionization fractions, most noticeably in the mass-weighted case, attributable to the fact that the regions of slower-moving gas (1-10 km/s) on the outskirts of the outflow are denser than the fast-moving jet cavity.

Figure~\ref{fig:chiprofiles_ratio} also shows observed ionization fractions in four observed knots in the outflowing jet of the massive protostar G35.2-0.74N \citep{Fedrianietal2019}. The reported values are $\sim 0.1$, with a slight decline as one goes from $\sim 3,000$ to 20,000~au. We note that these observations are based on measurements from regions that are emitting in the NIR via [FeII] emission lines, with this emission being used to derive the total number density of H nuclei. The electron density is derived independently from both the [FeII] emission and cm free-free emission from the knots, with each method giving similar results. The [FeII] line emission shows line of sight velocity shifts of $\sim 100\:{\rm km\:s}^{-1}$. Thus we consider that the most relevant comparison between our results and the G35.2-0.74N data is for mass-averaged results for the case with outflow velocity $\geq100\:{\rm km\:s}^{-1}$. For this case, we see that our model ionization fractions can begin to match the observed data near 20,000~au when $m_*\gtrsim 12\:M_\odot$. To match the inner knots requires somewhat higher protostellar masses. However, these results are sensitive to our choice of cooling model, i.e., Case B compared to Case A (see Fig.~\ref{fig:chiprofiles_noratio}). The protostellar mass of G35.2-0.74N is quite uncertain, but from spectral energy distribution (SED) fitting, \citet{2023ApJ...942....7F} estimate $m_*=19^{+9}_{-6}M_\odot$. Overall, we conclude that our simulation results for ionization fraction can match the observed values of G35.2-0.74N as long as $m_*\gtrsim 12\:M_\odot$, although the profile along the jet and its detailed structure, i.e., presence of knots, show differences and/or are quite sensitive to modeling choices for the treatment of cooling and choice of velocity threshold to define the outflow material.

\begin{deluxetable*}{cc|ccc|ccc|ccc}
\tablecaption{Average Ionization Fractions for Case B (with cooling), corresponding to the data in Figs.\ \ref{fig:chimaps_4000_ratio} and \ref{fig:chimaps_25000_ratio}.\label{tab:avg_chi_ratio}} 
\tablehead{\colhead{Scale} & 
\colhead{$m_*$} & \multicolumn{3}{c}{Mass-Weighted} & \multicolumn{3}{c}{Volume-Weighted} & \multicolumn{3}{c}{Emission-Weighted} \\\colhead{[au]} &
 \colhead{[$M_\odot$]} & \colhead{$\geq$1 km/s} & \colhead{$\geq$10 km/s} & \colhead{$\geq$100 km/s} & \colhead{$\geq$1 km/s} & \colhead{$\geq$10 km/s} & \colhead{$\geq$100 km/s} & \colhead{$\geq$1 km/s} & \colhead{$\geq$10 km/s} & \colhead{$\geq$100 km/s}
 }
 \startdata
	&	1.4	&	2.86e-06	&	1.86e-05	&	5.76e-01	&	1.97e-05	&	2.17e-05	&	5.76e-01	&	7.97e-05	&	2.66e-05	&	6.07e-01\\
	&	2.0	&	9.75e-06	&	5.57e-05	&	5.78e-01	&	4.38e-05	&	6.23e-05	&	5.78e-01	&	1.97e-04	&	7.59e-05	&	5.92e-01\\
	&	4.0	&	6.81e-06	&	6.98e-06	&	5.59e-01	&	1.34e-05	&	7.88e-06	&	5.59e-01	&	3.16e-05	&	1.04e-05	&	5.39e-01\\
4000	&	8.0	&	5.21e-05	&	6.00e-05	&	5.81e-01	&	7.52e-05	&	6.78e-05	&	5.81e-01	&	1.85e-04	&	8.79e-05	&	5.60e-01\\
	&	12.0	&	1.08e-04	&	1.11e-03	&	5.79e-01	&	5.01e-04	&	1.65e-03	&	5.81e-01	&	2.34e-03	&	2.16e-03	&	6.37e-01\\
	&	16.0	&	2.30e-04	&	5.17e-03	&	6.24e-01	&	1.12e-03	&	6.69e-03	&	6.25e-01	&	6.41e-03	&	7.49e-03	&	6.74e-01\\
	&	24.0	&	1.39e-04	&	6.08e-03	&	6.41e-01	&	1.26e-03	&	8.94e-03	&	6.45e-01	&	7.27e-03	&	1.06e-02	&	7.30e-01\\
\hline
	&	1.4	&	1.10e-05	&	3.64e-03	&	7.83e-01	&	3.18e-05	&	4.07e-03	&	7.83e-01	&	2.90e-04	&	6.17e-03	&	8.90e-01\\
	&	2.0	&	1.25e-05	&	2.67e-03	&	7.13e-01	&	6.93e-05	&	3.33e-03	&	7.13e-01	&	4.50e-04	&	5.36e-03	&	8.37e-01\\
	&	4.0	&	3.55e-06	&	2.36e-04	&	7.86e-01	&	3.49e-05	&	4.09e-04	&	7.86e-01	&	1.13e-04	&	5.43e-04	&	8.14e-01\\
25000	&	8.0	&	2.03e-05	&	2.28e-03	&	6.79e-01	&	9.99e-05	&	4.11e-03	&	6.79e-01	&	5.96e-04	&	5.41e-03	&	7.92e-01\\
	&	12.0	&	7.55e-05	&	1.23e-02	&	6.01e-01	&	7.43e-04	&	1.92e-02	&	6.02e-01	&	5.09e-03	&	2.49e-02	&	6.74e-01\\
	&	16.0	&	3.41e-04	&	3.73e-02	&	6.36e-01	&	2.81e-03	&	5.52e-02	&	6.36e-01	&	2.23e-02	&	6.59e-02	&	6.82e-01\\
	&	24.0	&	4.35e-04	&	5.45e-02	&	6.45e-01	&	3.26e-03	&	7.34e-02	&	6.46e-01	&	2.32e-02	&	8.78e-02	&	7.29e-01\\
\hline
\enddata
\end{deluxetable*}




\subsection{Free-Free Radio Emission}
\subsubsection{Intensity Maps}
\label{sec:results_intensity}
\begin{figure*}
    \centering
    \includegraphics[width=1.0\linewidth]{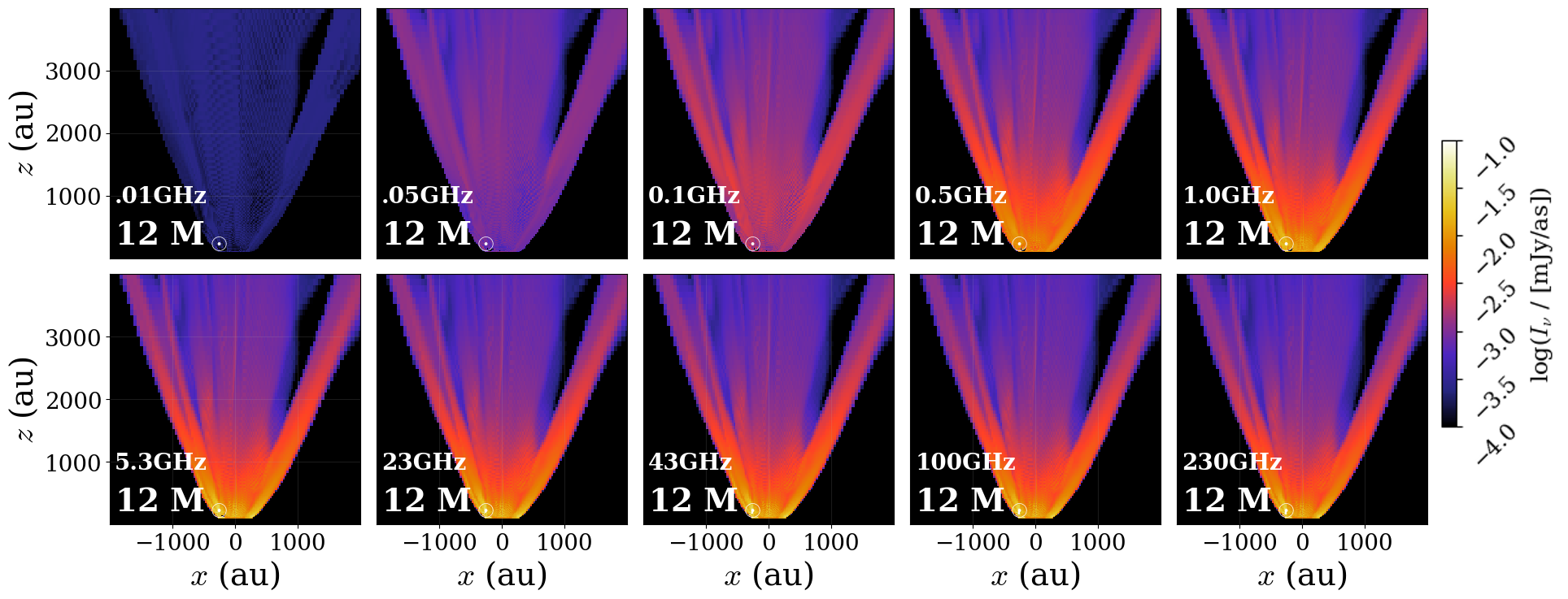}
    \includegraphics[width=1.0\linewidth]{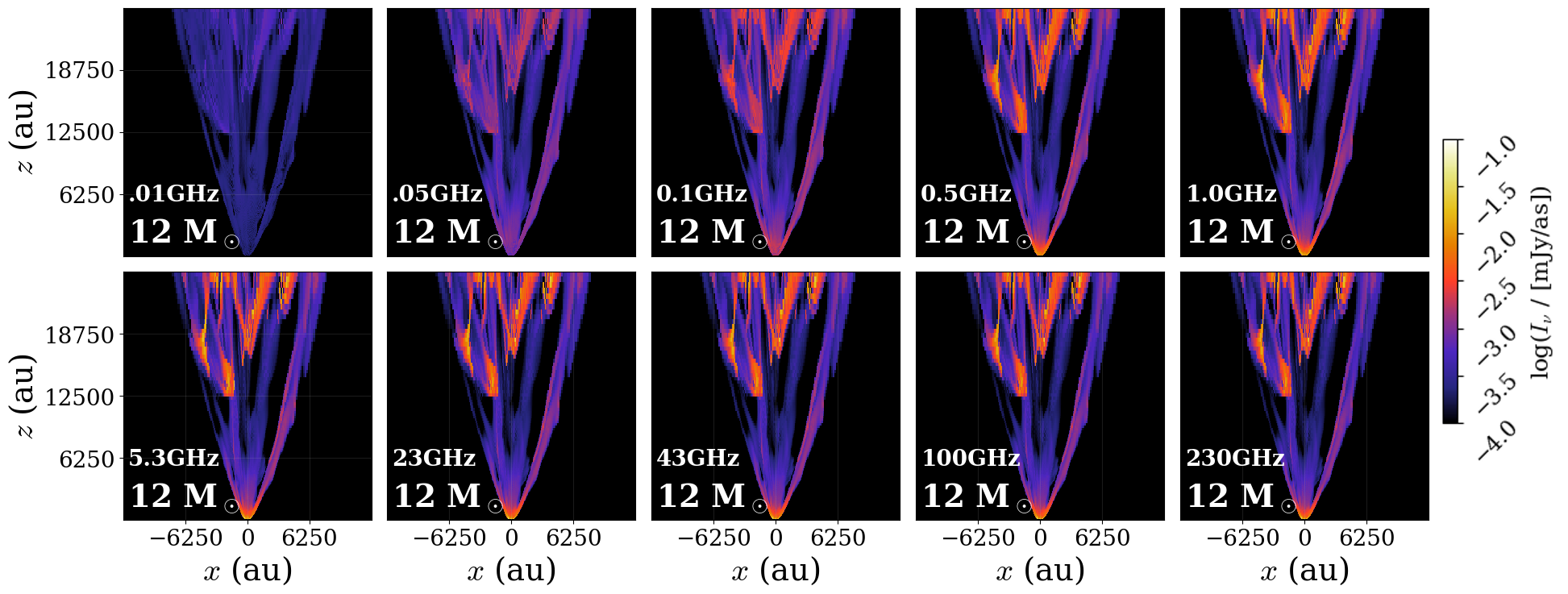}
    \caption{Free-free intensity maps for Case B (with cooling) for the $12\:M_\odot$  (54,000 yr) snapshot at 0.01, 0.05, 0.1, 0.5, 1, 5.3, 23, 43, 100, and 230 GHz, over the ranges $-2000\:\mathrm{au}<x<2000\:\mathrm{au}$, $0\:\mathrm{au}<z<4000\:\mathrm{au}$ (\textit{top}) and $-12500\:\mathrm{au}<x<12500\:\mathrm{au}$, $0\:\mathrm{au}<z<25000\:\mathrm{au}$ (\textit{bottom}). 
    \label{fig:intensitymaps_12M_ratio}}
\end{figure*}

The free-free intensities at frequencies of 0.01, 0.05, 0.1, 0.5, 1, 5.3, 23, 43, 100, and 230 GHz were calculated for both Case A (no cooling) and Case B (with cooling).
Examples are shown for the $m_*=12\:M_\odot$ (54,000~yr) snapshot in Fig.~\ref{fig:intensitymaps_12M_ratio} (see Fig.~\ref{fig:intensitymaps_12M_noratio} for the equivalent Case A results). Although the morphology appears similar at each frequency, it is dimmest at 0.01 GHz, brightens significantly up to 0.5 GHz, then maintains similar brightness up to 230 GHz. A quantitative analysis of the radio spectral energy distribution is provided below in \S\ref{sec:results_spectra}.

\begin{figure*}
    \centering
    \includegraphics[width=0.49\linewidth]{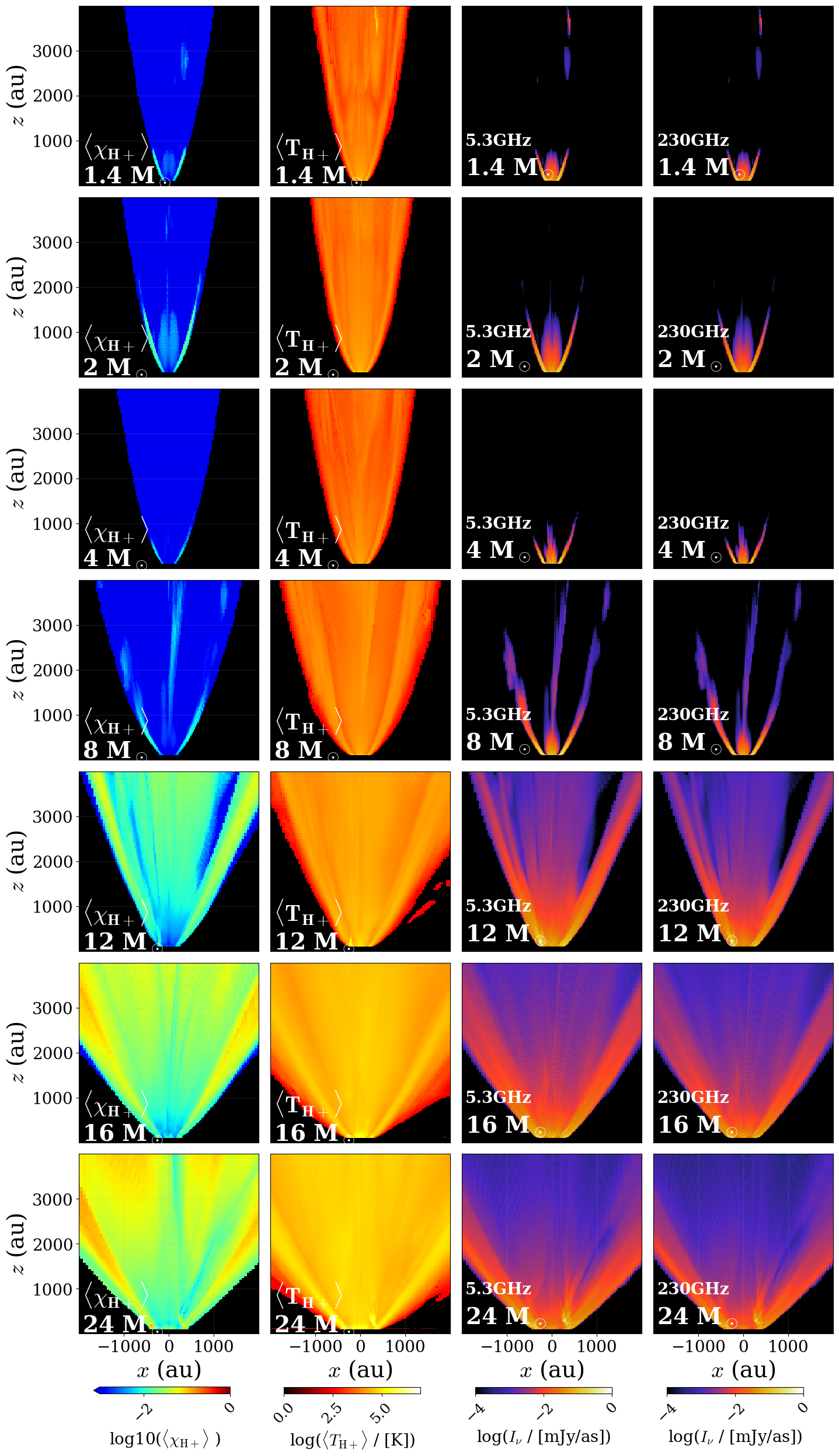}     
    \includegraphics[width=0.49\linewidth]{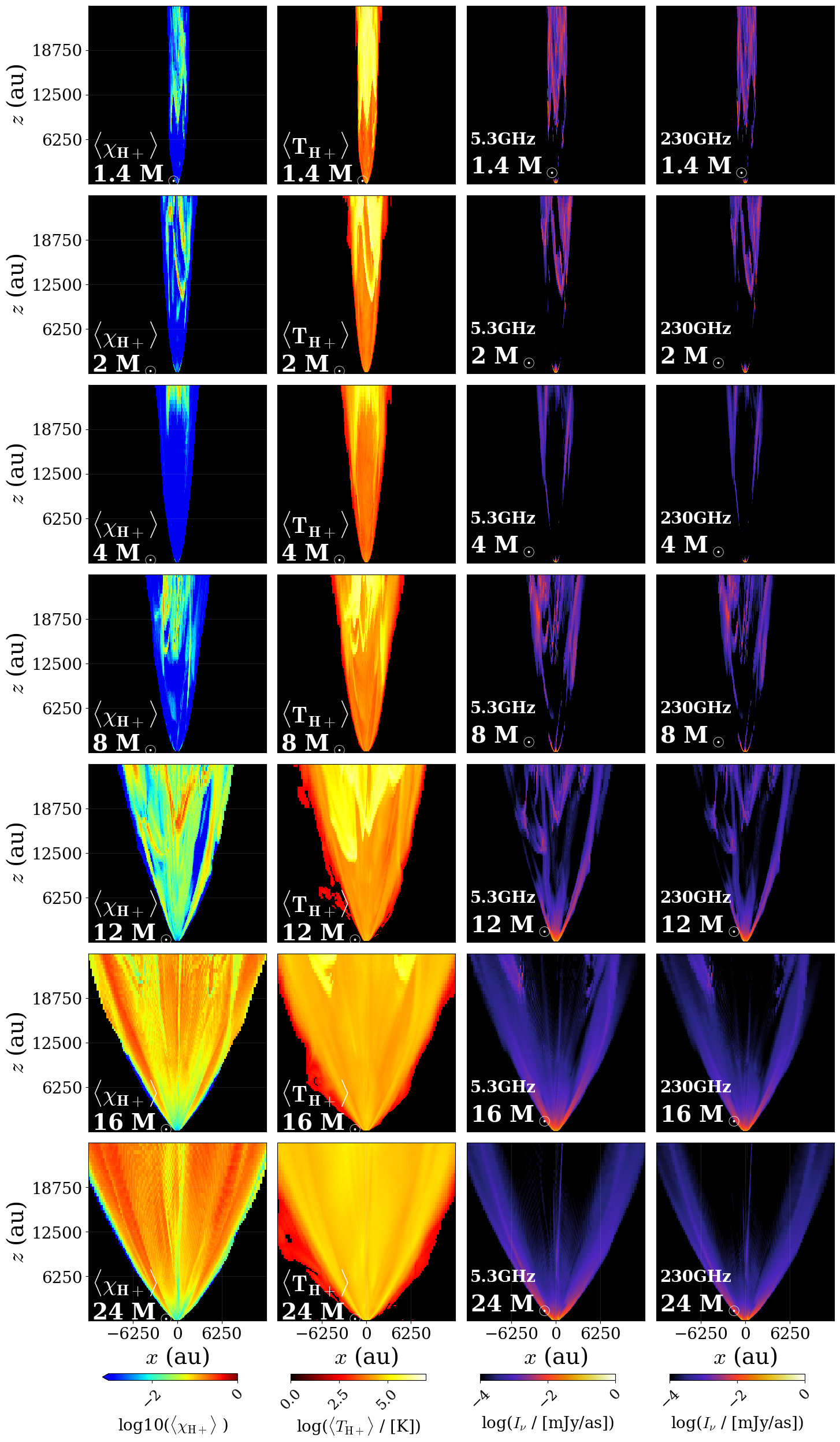}  
    \caption{Time evolution of shock-emission variables projected along the y-direction for Case B (with cooling). \textit{Left set:} Displayed domain is for inner 4,000~au scale.
    \textit{Right set:} Displayed domain is for global 25,000~au scale. 
    Within each set, from left to right, the columns show mass-weighted average ionization fraction of the jet with a velocity cutoff of 100 km s$^{-1}$, ionized-mass-weighted temperature, 5.3 GHz intensity, and 230 GHz intensity.
    \label{fig:projections_ratio}}
\end{figure*}

We also examine the time evolution of these intensity maps as the protostar grows in mass. 
Fig.\ \ref{fig:projections_ratio} illustrates the evolution of the 5.3~GHz and 230~GHz intensity maps, along with mass-weighted, $v_\mathrm{cutoff} = 100$~km/s ionization fraction projections and ionized mass-weighted temperature projections. Fig.\ \ref{fig:projections_noratio} shows the equivalent Case A results.
The side-by-side intensity and projection maps show that the regions of highest intensity and ionization correspond to average ionized-gas temperatures of $\sim 10^7$ K. The maps also show a gap in both 5.3~GHz and 230~GHz emissions corresponding to the low-shock activity region of $1000 \mathrm{\ au}\lesssim z \lesssim 5000 \mathrm{\ au}$ for snapshots up to 39,000 years, identifiable in Fig. \ref{fig:slices} as having temperatures $\lesssim10^4\:$K and little to no emissivity.  


\begin{figure}
    \centering
    \includegraphics[width=\columnwidth]{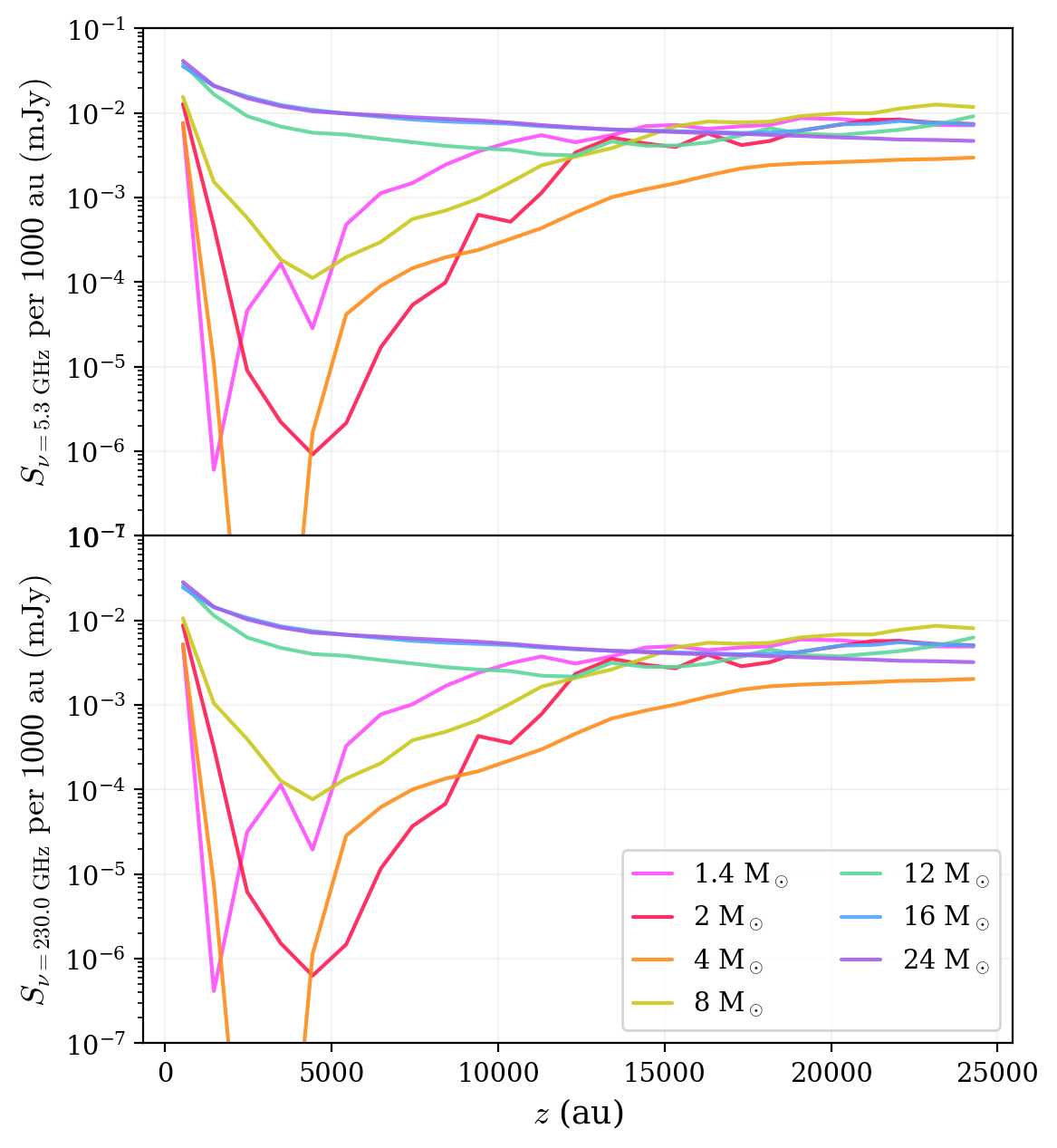}
    \caption{Free-free emission fluxes for Case B (with cooling) at 5.3 GHz (top panel) and 230 GHz (bottom panel) as a function of height; fluxes are calculated by integrating the intensity $I_\nu$ along the $y$-direction over 1000~au wide regions covering $-25000\:\mathrm{au}<x<25000\:\mathrm{au}$, $h-500\:\mathrm{au} < z < h+500\:\mathrm{au}$ for each height, $h$. Different colors represent the simulation at increasing protostellar masses: 1.4, 2, 4, 8, 12, 16, and 24~$M_\odot$ are shown in pink, red, orange, yellow, green, blue, and purple, respectively.}
    \label{fig:fluxprofileB}
\end{figure}

Profiles of the 5.3 and 230 GHz radio emission as a function of height were calculated by integrating the flux over 1,000 au wide regions in the $z$-direction, as shown in Fig.~\ref{fig:fluxprofileB} (see Fig. \ref{fig:fluxprofileA} for Case A).
These profiles clearly show the low radio flux region that is present around several thousand au in the early evolutionary stages ($m_*\lesssim 8\:M_\odot$). This stage is associated with a relatively collimated outflow cavity, so that there is limited emission from shocks of the outflowing gas with the infall envelope at these locations. However, at later stages, when the opening angle becomes wider, then there are stronger shocks present at these boundaries.


\subsubsection{Integrated Fluxes and Spectra}
\label{sec:results_spectra}
Integrated flux values at 5.3 GHz and 230~GHz were calculated for each evolutionary stage, i.e., $m_*=1.4, 2, 4, 8, 12, 16, 24\:M_\odot$ including flux within scales of $r$ = 500, 1000, 2000, 4000, 8000, 16000, and 25000 au (doubled to account for the bipolar nature of the outflow). For this analysis, at each stage an average was made of 11 consecutive snapshots, each 10 years apart, covering the 100~yr period after each mass was achieved in the simulation. To assess the short timescale variability, we also measured the standard deviation of the values of ${\rm log}\:S_\nu$ (with these results discussed below in \S\ref{sec:results_variability}). The values of the fluxes and their standard deviations are presented in Table \ref{tab:fluxes_caseB} for Case B (see Table \ref{tab:fluxes_caseA} for Case A). 



\begin{figure*}
    \centering
    \includegraphics[width=\linewidth]{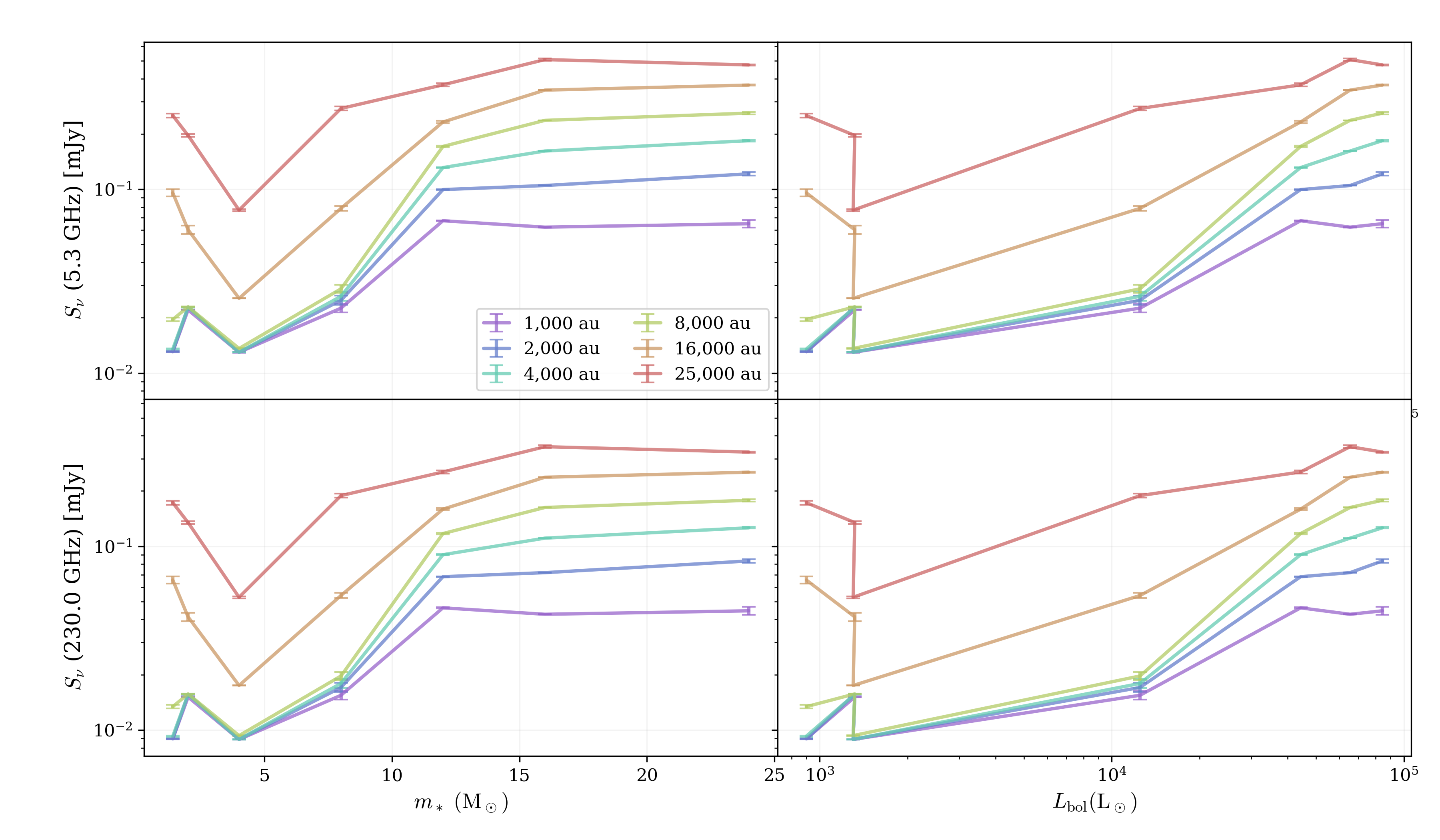}
    \caption{Average 5.3 GHz (\textit{top row}) and 230 GHz (\textit{bottom row}) fluxes for Case B, as a function of mass (\textit{left column}) and bolometric luminosity (\textit{right column}), integrated over regions of $-r/2 < x < r/2$, $0<z<r$ for $r$ = 1000, 2000, 4000, 8000, 16000, and 25000 au (as labeled; see Table~\ref{tab:fluxes_caseB}), then doubled to account for the counter jet.\label{fig:fluxvsmass_caseB}}
\end{figure*}

The evolution of the integrated fluxes as a function of increasing mass and bolometric luminosity is shown in Fig.\ \ref{fig:fluxvsmass_caseB}.
We see that the fluxes generally increase as $m_*$ grows, although there is a local minimum at $m_*\sim 4\:M_\odot$, which is related to a stage of protostellar evolution when the star is relatively swollen so that its outflows are relatively slow \citep{Zhangetal2014_paperIII}. At higher protostellar masses, the radio fluxes are seen to reach near constant values, although one should note that this is a regime where the photoionization process, which is not included in our modeling, will begin to play a more important role \citep{Tanakaetal2016}.

\begin{deluxetable*}{cccccccccc}
\tabletypesize{\footnotesize}
\tablecaption{Radio Free-Free Fluxes at 5.3~GHz and 230~GHz for Case B, i.e., with Cooling \label{tab:fluxes_caseB}} 
\tablehead{
\colhead{$m_*$} & \colhead{$t$} &  \colhead{$L_{\rm bol}$}  &  \multicolumn{6}{c}{log ($S_\nu$/[mJy])} \\
 \colhead{ [$M_\odot$]} & \colhead{[yr]} & \colhead{[$L_\odot$]} & \colhead{500~au} & \colhead{1,000~au} & \colhead{2,000~au} & \colhead{4,000~au} & \colhead{8,000~au} & \colhead{16,000~au} & \colhead{25,000~au}
 }
\startdata
\multicolumn{10}{c}{5.3 GHz} \\
1.4	&	4000	&	899.98	&	-1.917$\pm$0.002	&	-1.884$\pm$0.003	&	-1.884$\pm$0.003	&	-1.869$\pm$0.003	&	-1.709$\pm$0.010	&	-1.020$\pm$0.020	&	-0.599$\pm$0.011	\\
2.0	&	9000	&	1317.53	&	-1.778$\pm$0.001	&	-1.657$\pm$0.001	&	-1.642$\pm$0.002	&	-1.642$\pm$0.002	&	-1.640$\pm$0.002	&	-1.220$\pm$0.023	&	-0.707$\pm$0.008	\\
4.0	&	21000	&	1301.15	&	-1.912$\pm$0.002	&	-1.887$\pm$0.002	&	-1.887$\pm$0.002	&	-1.887$\pm$0.002	&	-1.867$\pm$0.002	&	-1.594$\pm$0.001	&	-1.113$\pm$0.005	\\
8.0	&	39000	&	12499.0	&	-1.707$\pm$0.023	&	-1.647$\pm$0.024	&	-1.604$\pm$0.025	&	-1.582$\pm$0.025	&	-1.542$\pm$0.022	&	-1.104$\pm$0.014	&	-0.560$\pm$0.010	\\
12.0	&	54000	&	44323.6	&	-1.407$\pm$0.004	&	-1.171$\pm$0.003	&	-1.002$\pm$0.002	&	-0.882$\pm$0.003	&	-0.767$\pm$0.004	&	-0.634$\pm$0.006	&	-0.431$\pm$0.008	\\
16.0	&	68000	&	65461.5	&	-1.495$\pm$0.002	&	-1.206$\pm$0.001	&	-0.979$\pm$0.001	&	-0.792$\pm$0.001	&	-0.625$\pm$0.001	&	-0.460$\pm$0.000	&	-0.294$\pm$0.007	\\
24.0	&	94000	&	84459.8	&	-1.622$\pm$0.003	&	-1.187$\pm$0.021	&	-0.916$\pm$0.010	&	-0.735$\pm$0.004	&	-0.586$\pm$0.007	&	-0.433$\pm$0.004	&	-0.323$\pm$0.003	\\
\hline
\multicolumn{10}{c}{230.0 GHz} \\
1.4	&	4000	&	899.98	&	-2.081$\pm$0.002	&	-2.048$\pm$0.003	&	-2.048$\pm$0.003	&	-2.033$\pm$0.003	&	-1.873$\pm$0.010	&	-1.184$\pm$0.020	&	-0.763$\pm$0.011	\\
2.0	&	9000	&	1317.53	&	-1.942$\pm$0.001	&	-1.821$\pm$0.001	&	-1.806$\pm$0.002	&	-1.806$\pm$0.002	&	-1.804$\pm$0.002	&	-1.384$\pm$0.023	&	-0.871$\pm$0.008	\\
4.0	&	21000	&	1301.15	&	-2.076$\pm$0.002	&	-2.051$\pm$0.002	&	-2.051$\pm$0.002	&	-2.051$\pm$0.002	&	-2.031$\pm$0.002	&	-1.758$\pm$0.001	&	-1.277$\pm$0.005	\\
8.0	&	39000	&	12499.0	&	-1.871$\pm$0.023	&	-1.811$\pm$0.024	&	-1.768$\pm$0.025	&	-1.746$\pm$0.025	&	-1.706$\pm$0.022	&	-1.268$\pm$0.014	&	-0.724$\pm$0.010	\\
12.0	&	54000	&	44323.6	&	-1.571$\pm$0.004	&	-1.335$\pm$0.003	&	-1.165$\pm$0.002	&	-1.046$\pm$0.003	&	-0.931$\pm$0.004	&	-0.798$\pm$0.006	&	-0.595$\pm$0.008	\\
16.0	&	68000	&	65461.5	&	-1.659$\pm$0.002	&	-1.370$\pm$0.001	&	-1.143$\pm$0.001	&	-0.956$\pm$0.001	&	-0.789$\pm$0.001	&	-0.624$\pm$0.000	&	-0.458$\pm$0.007	\\
24.0	&	94000	&	84459.8	&	-1.786$\pm$0.003	&	-1.351$\pm$0.021	&	-1.080$\pm$0.010	&	-0.899$\pm$0.004	&	-0.750$\pm$0.007	&	-0.597$\pm$0.004	&	-0.487$\pm$0.003	\\
\enddata
\tablecomments{5.3~GHz and 230~GHz fluxes are integrated over $-r/2<x<r/2$, $0<z<r$ for each scale r = 500, 1000, 2000, 4000, 8000, 16000, and 25000 au, multiplied by 2 to account for the counter jet. The exception is that for 25000~au, the integration is carried out over the whole simulation domain, which extends over $-15000\ \mathrm{au} <x<15000\ \mathrm{au}$, $0\ \mathrm{au}<z<25000\ \mathrm{au}$.}
\end{deluxetable*}

\begin{figure*}
    \centering
    \includegraphics[width=1.0\textwidth]{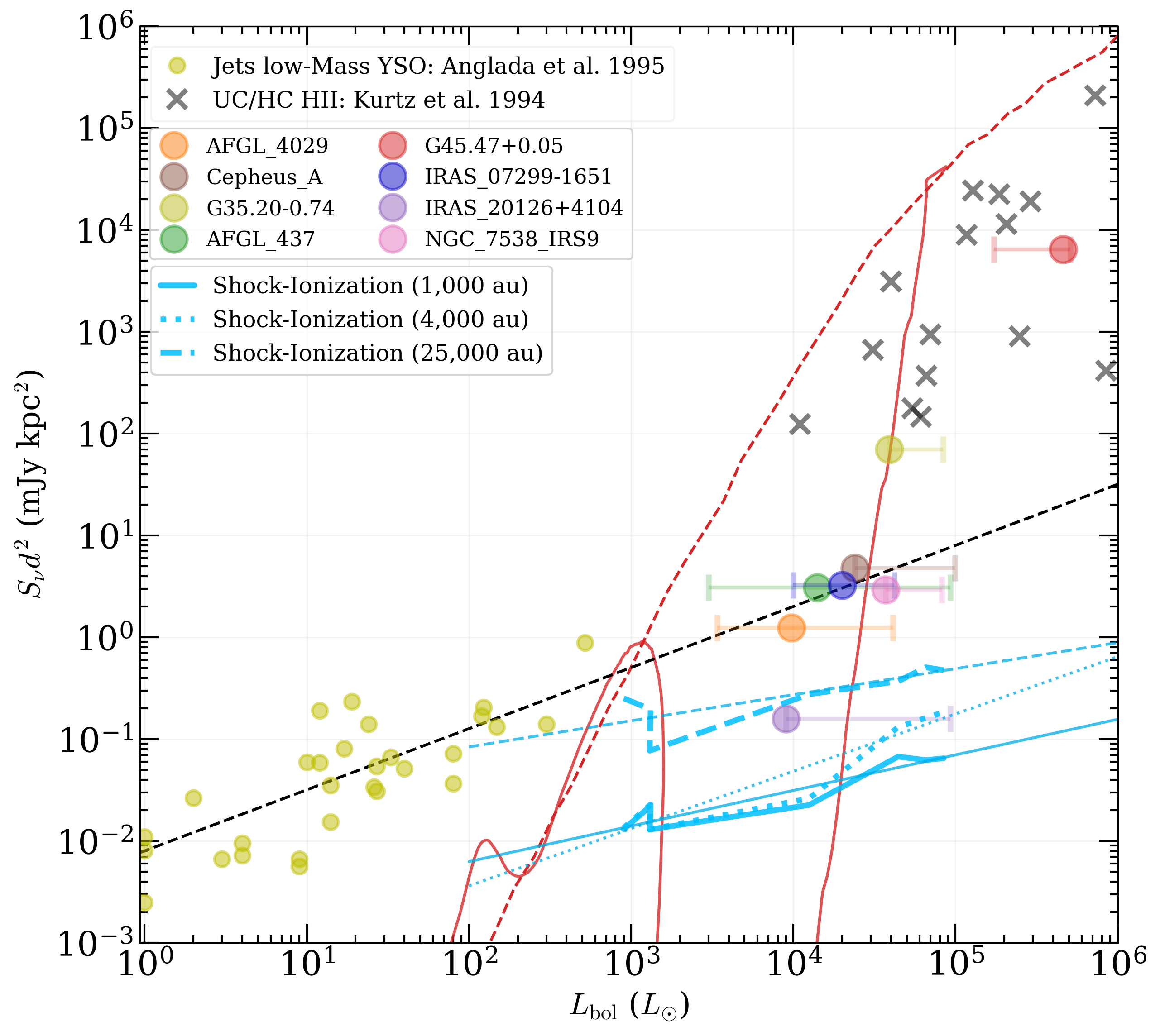}
    \caption{
    Radio continuum luminosity metric ($S_\nu d^2$) at 5.3 GHz versus bolometric luminosity. The solid, dotted and dashed lines correspond to the evolution in fluxes from 1.4 to 25~$M_\odot$ on scales with $r=1000, 4000, 25000\:$au where the integration is carried out over $-r/2 < x < r/2$, $0 < z < r$ and then doubled to account for the counter jet (see Table~\ref{tab:fluxes_caseB}).
    Power law fits are shown by the thin lines of the corresponding line style.
    The red dashed line shows radio emission from an optically thin HII region photoionized by zero-age main sequence stars \citep{Thompson1984}, while the red solid line is the equivalent model for the fiducial TCA model protostar (initial core mass $M_c=60\:M_\odot$; mass surface density of clump environment $\Sigma_{\rm cl}=1\:{\rm g\:cm}^{-2}$) \citep{Tanakaetal2016}.
    Yellow circles show low-mass protostar \citep{Anglada1995} with a power law fit of $8\times 10^3(L_\mathrm{bol})^{0.6}$ (black dashed line) \citep{Angladaetal2015}. 
    Large circles are eight protostars from the SOFIA Massive Star Formation Survey  \citep{DeBuizeretal2017_SOMAI,Roseroetal2019} (see legend).
    Crosses are ultracompact HII regions \citep{Kurtzetal1994}.     
    \label{fig:AngladaDiagram}}
\end{figure*}


Figure~\ref{fig:AngladaDiagram} presents the 5.3~GHz radio luminosity (measured by the metric $S_\nu d^2$) versus bolometric luminosity diagram. Our shock-ionization (Case B) models integrated on scales of 1000, 4000, and 25000~au are shown, along with power law fits $S_\nu d^2 \propto L_{\rm bol}^{\gamma}$ with $\gamma = 0.349, 0.563, 0.256$, respectively. The semi-analytic photoionization model of \citet{Tanakaetal2016} for the fiducial TCA model, i.e., the same used in our numerical simulations, is shown by the red solid line. We see that in the luminosity range $L_{\rm bol}\lesssim 2 \times 10^4\:L_\odot$ the contribution from photoionization is generally expected to be insignificant compared to shock ionization, with a minor exception near $L_{\rm bol}\sim 10^3\:L_\odot$. This limited contribution of photoionization is due to the typically low photospheric temperatures and relatively low luminosities of the TCA protostellar models for $m_*\lesssim 10\:M_\odot$, i.e., before Kelvin-Helmholtz contraction towards the zero age main sequence (ZAMS) (Fig.~\ref{fig:AngladaDiagram} also shows a reference model of photoionization from ZAMS stars). 
We also note that the predictions of the photoionization contribution at early stages are very uncertain since in this regime the accretion luminosity dominates $L_{\rm bol}$ and the effective temperature of its emission depends on assumptions about the scale of the region where it is liberated: the \citet{Tanakaetal2016} model assumes the protostar's internal and boundary layer accretion luminosity are released together from the stellar surface with a single effective photospheric temperature.

Figure~\ref{fig:AngladaDiagram} also shows observational data from several samples of protostars. First, a sample of low-mass, lower-luminosity ($\lesssim10^3\:L_\odot$) protostars have been studied by \citet{Anglada1995} and \citet{Angladaetal2015}, including a power law fit $(S_\nu d^2 / {\rm [mJy\:kpc^2]}) = 8\times 10^3 (L_{\rm bol}/L_\odot)^{0.6}$. These sources are expected to be powered by shock ionization. We note that an extrapolation of this power law to higher bolometric luminosities predicts radio luminosities that are factors of several to ten times larger than our models of the entire 25,000~au domain. However, it is uncertain whether this extrapolation of this purely empirical relation is valid. 

The next data set is a sample of eight relatively massive protostars from the SOFIA Massive (SOMA) Star Formation Survey \citep{DeBuizeretal2017_SOMAI}. These have had their 5.3~GHz fluxes measured on various scales \citep{Roseroetal2019}. The values shown in Fig.~\ref{fig:AngladaDiagram} are evaluated from the ``Intermediate'' scale, which is designed to capture any resolved radio jet (when there is no clear evidence for a resolved jet, the ``Inner'' scale is used). We see most of these protostellar sources have radio luminosities about 10 times greater than our Case B 25,000~au scale model. IRAS~20126+4104 sits close to this model, while G35.20-0.74N and G45.47+0.05 are about $10^3$ to $10^4$ times more luminous than the shock-ionization model. A sample of ultracompact (UC) HII regions \citep{Kurtzetal1994} have similarly elevated radio luminosities. 

For G45.47+0.05 and many of the UC HII regions, which have high bolometric luminosities, $\gtrsim 10^5\:L_\odot$, photoionization is expected to be dominant. For the lower luminosity sources, it is generally unclear whether their radio emission is dominated by shock ionization or photoionization. However, in the case of G35.20-0.74N, \citet{Fedrianietal2019} argued that the modest ionization fractions of $\sim 0.1$ indicate a preference for the shock ionization mechanism, at least in the jet knots. In this case, we see that our fiducial Case B models underpredict the required radio flux by a factor of $\sim 10^3$. We show in Appendix~\ref{app:caseA} that the Case A model, i.e., without a cooling limit, boosts the radio luminosities from the inner 1,000 and 4,000~au regions to $\sim 1$ to 10~mJy~$\rm kpc^2$ levels, while the 25,000~au scale can reach the 100~mJy~$\rm kpc^2$ level that is relevant to G35.20-0.74.

\begin{figure}
    \centering
    \includegraphics[width=\columnwidth]{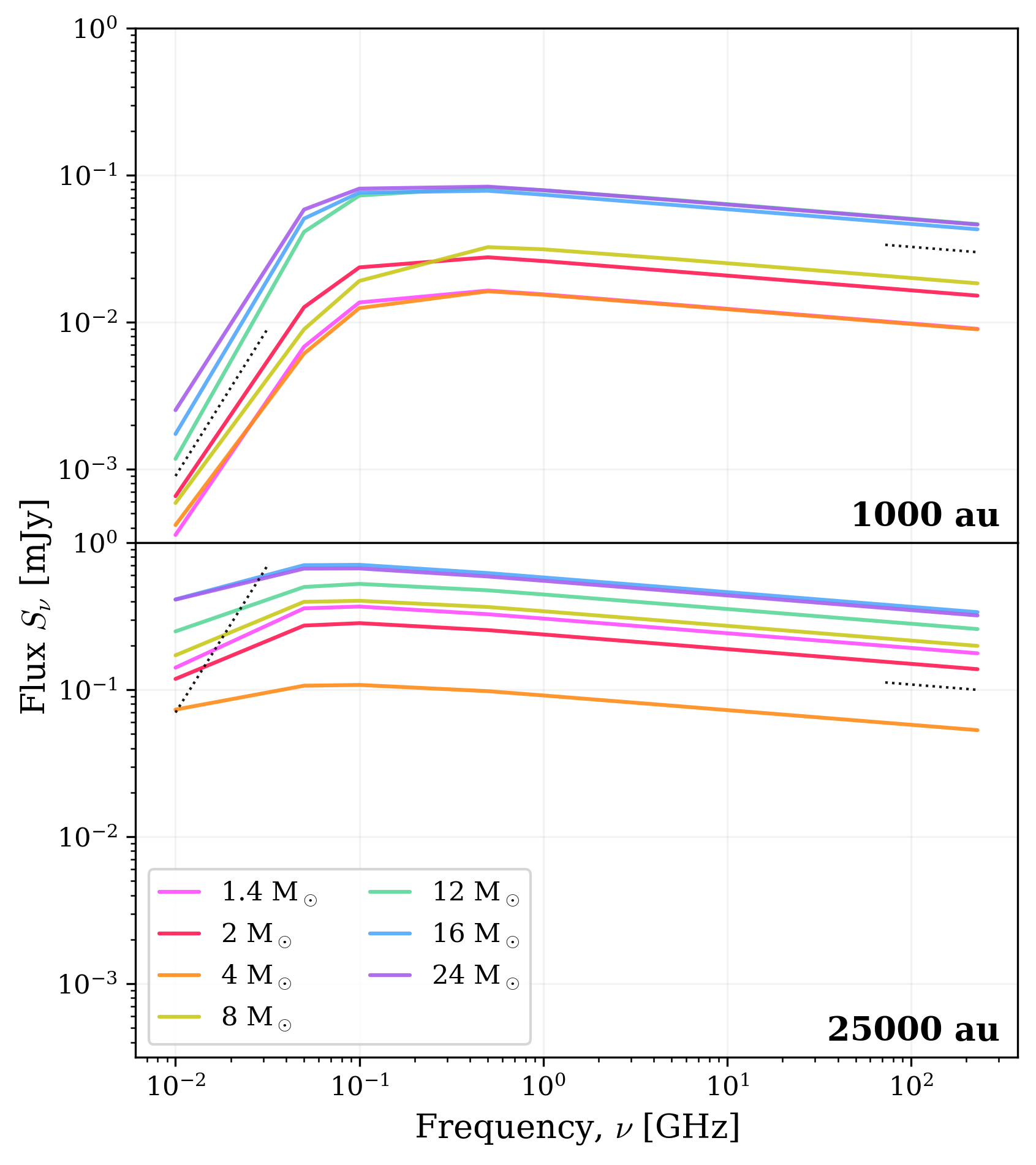}
    \caption{Radio free-free emission spectra of the shock-ionized gas (Case B) in the 1,000 au region (\textit{top}) and the full domain (\textit{bottom}) for the different protostellar evolutionary stages (as labeled). 
    Dotted line segments show the spectral indices of optically thick conditions ($S_\nu\propto \nu^2$) that are expected at low-frequencies and of optically thin condition ($S_\nu\propto \nu^{-0.1}$) that are expected at high frequencies.
    }\label{fig:spectralevolution_caseB}
\end{figure}

The full spectrum of a protostar's radio emission provides additional information with which to constrain the models. The overall radio free-free spectra of the shock-ionized protostellar outflows are shown in Fig.~\ref{fig:spectralevolution_caseB}, for both the inner ($\leq1,000\:$au) and global ($\leq25,000\:$au) scales.
The spectra of the inner region show the expected optically thick behavior at low frequencies, i.e., $S_\nu \propto \nu^2$, and then turnover at frequencies $\sim 0.1\:$GHz to the optically thin regime where $S_\nu \propto \nu^{-0.1}$. The spectra of the global region stay closer to the optically thin condition down to much lower frequencies, i.e., with flat or only slightly positive spectral indices down to 0.01~GHz.

We note that the radio spectral indices of G35.20-0.74N at the inner and intermediate scales are $\alpha_{\rm inner} = 0.7\pm 0.1$ and $\alpha_{\rm inter} = -0.2\pm 0.1$, respectively \citep{Roseroetal2019}. Such values are consistent with the general trends predicted by the shock ionization models shown in Fig.~\ref{fig:spectralevolution_caseB}, where the inner scale shows a higher degree of optical depth.

\subsubsection{Flux Variability} 
\label{sec:results_variability}
\begin{figure}
    \centering
    \includegraphics[width=\columnwidth]{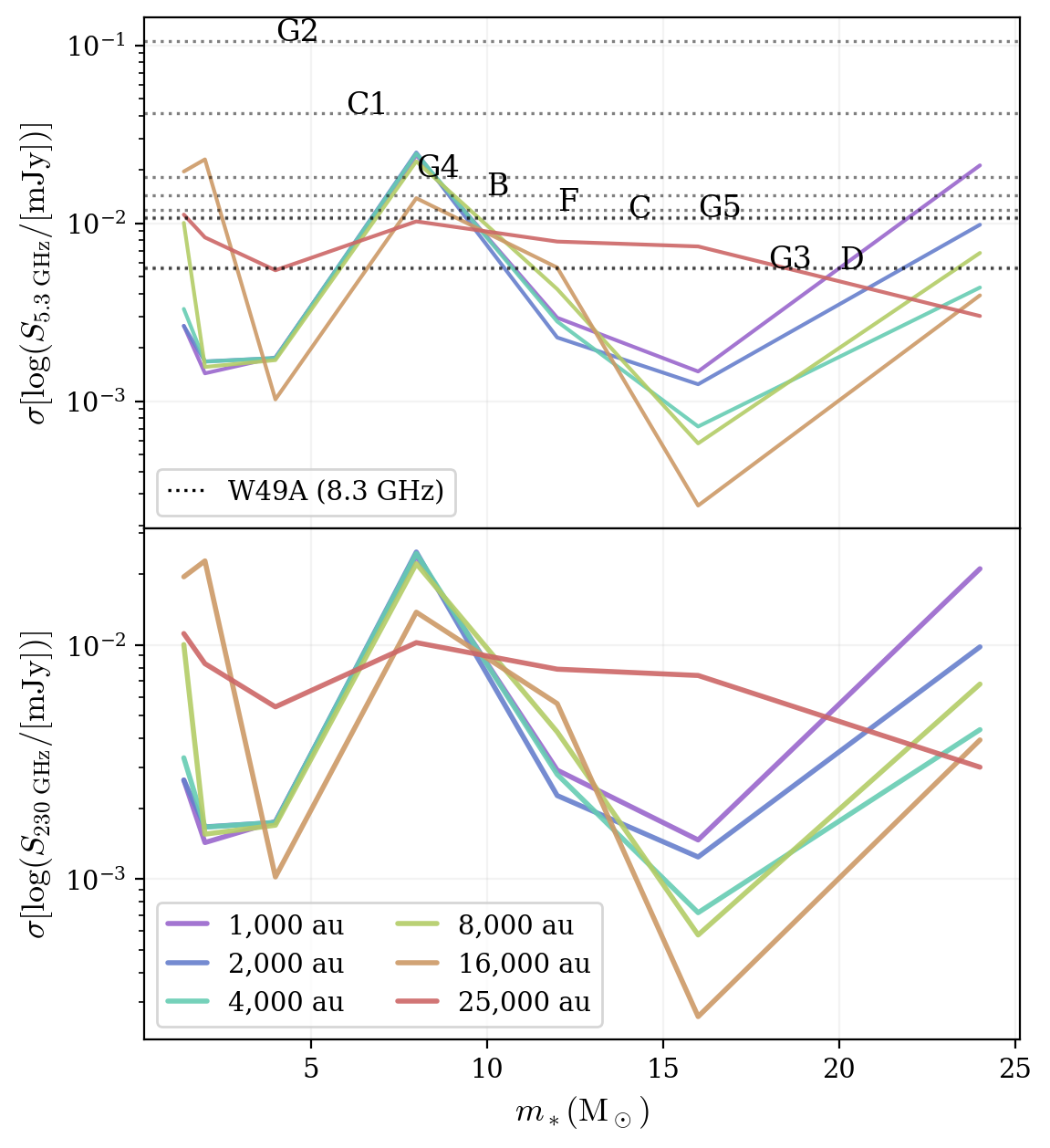}
    \caption{Flux variability, i.e., standard deviation of $\log S_\nu$ over 10-year intervals at 5.3 GHz (\textit{top panel}) and 230 GHz (\textit{bottom panel}) for 11 values spaced over 100 years, as a function of mass, for Case B.  
    The black dotted lines in the top panel show the 20-year standard deviations of the 8.3~GHz flux from nine sources in massive star-forming region W49A; these regions are labeled from highest to lowest variability: G2, C1, G4, B, F, C, G5, G3, and D \citep{dePreeetal2018}. 
    \label{fig:percentvariation_caseB}}
\end{figure}

The radio flux variability was evaluated for a sample of 11 snapshots of the simulation separated by 10-year periods, i.e., over a duration of 100 yr, at each fiducial evolutionary stage of the simulation, i.e., $m_*=1.4, 2, 4, 8, 12, 16, 24\:M_\odot$.
These flux variations are shown in Fig.~\ref{fig:percentvariation_caseB} (see Fig. \ref{fig:percentvariation_caseA} for Case A).

At both 5.3 and 230~GHz, the average 10-year variability is $\sigma({\rm log}\:S_\nu / [{\rm mJy}]) \sim 0.001$ to 0.02, i.e., variations in flux of $\sim 0.2$ to 5\%. As expected, variations tend to be larger on smaller scales. On detailed examination of the radio intensity images, we find most variability is due to motion of particular hot spots of emission in the outflowing gas. Note, that a characteristic flow crossing timescale is $t_{\rm outflow}\sim1000\:{\rm au} / 1000\:{\rm km\:s}^{-1} \sim 5\:$yr.

Figure~\ref{fig:percentvariation_caseB} also shows the variability level of nine example observed hypercompact HII regions in W49A, which have measured 8.3~GHz flux variations over $\sim 20\:$yr timescales, i.e., from 1994 to 2015 \citep{dePreeetal2018}.
Most of the sources show variations at the $\sim$ few percent level, which is quite comparable to the variability seen in our simulated outflows. However, the greatest change in flux identified by \citet{dePreeetal2018} was that of source W49A/G2, which decreased by 40\% from 1994 to 2015. We note that our model outflow has a smoothly varying outflow injection rate, while real sources may suffer some short timescale variation in accretion/outflow activity. We consider that such accretion/outflow variability is likely to help generate greater levels of radio flux variability, which is a subject worthy of future study.

\section{Discussion and Conclusions}
\label{sec:discussion}

Shock ionization is expected to be the dominant source of ionized gas and associated radio emission in early-stage massive protostars. We have thus developed a model for shock-ionization of the disk wind driven outflow during an evolutionary sequence of massive star formation. In particular, the model has been applied as a post-processing step to the MHD simulations of \citet{Staffetal2023} of a protostar forming from an initial $60\:M_\odot$ core embedded in a $\Sigma_{\rm cl}=1\:{\rm g\:cm}^{-2}$ environment, i.e., the fiducial TCA model of massive star formation. The method involves estimating the post-shock conditions caused by converging flows between discrete cells in the simulation and then assessing the fraction of the cell that is filled based on the ratio of cooling to flow crossing time. Collisional ionization occurs in the shock heated gas. The free-free emission from this plasma was then calculated to predict fluxes and spectra in the radio regime.

The presented model is relatively simple and there are several caveats and limitations of the work. The first caveat is that the model involves post-processing, rather than a fully self-consistent MHD simulation with the full heating, cooling, and ionization processes. However, such numerical simulations would require very high resolution to be able to resolve the shock structures, highlighting the advantage and need of our subgrid model for the shocks. Nevertheless, there is still the question of the effect of numerical resolution on our results. To investigate this we have carried out ``low'' resolution simulations that have cells twice the size of those in our fiducial runs. We present their results for the radio spectra from the models in Appendix~\ref{app:res}. The overall result of increasing the resolution by a factor of two is typically to change the radio flux by within a factor of 2, although moderately larger differences can arise on certain scales and at certain frequencies (see Appendix~\ref{app:res}).

Another caveat is that the model is sensitive to how the cooling is treated for the post-shock gas. The simplest model with no cooling leads to much greater amounts of ionized gas and thus brighter radio emission, especially at higher frequencies where the structures are optically thin. The simplicity of our cooling-limited model, i.e., that sets the fraction of cells filled with post-shock gas to be $\sim t_{\rm cool}/t_{\rm flow}$, is such that one naturally expects systematic uncertainties in the quantitative model results at the level of a factor of a few. 

As already mentioned, the model does not include any contribution to ionization from photoionization. Thus the model predictions should be viewed as lower limits, which we expect to be superseded when protostellar masses become large. Evaluating the contribution from photoionization as an additional postprocessing step is deferred to a future paper in this series.

Finally, when comparing to observed sources one must remember that the model has been applied only to a single evolutionary track in TCA model parameter space of inital core mass and clump environment mass surface density. Other parts of this parameter space involve protostars accreting at different rates and with different stellar radii, leading to variations in the outflow speeds, densities and thus shock conditions. Exploring a wider range of this parameter space is also deferred to a future study.


The main conclusions of our study are the following:

\begin{enumerate}

    \item Shocks in a disk wind driven outflow, especially at its interface with the ambient infall envelope, produce localized regions that are heated to high temperatures, up to $\sim10^5$ to $10^7\:$K, with the maximum temperature reached increasing with increasing protostellar mass.

    \item The resulting ionization fractions of the outflowing gas quickly reach levels of $\sim0.1$ when considering volume-weighted averages. However, mass-weighted averages only reach such levels when $m_*\gtrsim 16\:M_\odot$ and when selecting outflowing gas via a high velocity threshold of $\sim100\:{\rm km\:s}^{-1}$. Yet, the ionization fraction of the material that contributes to the radio free-free emission (i.e., emission weighted) is close to fully ionized. The mass-weighted ionization fraction level of $\sim 0.1$ is similar to that inferred in jet knots of the massive protostar G35.20-0.74N by \citet{Fedrianietal2019}.

    \item In our cooling-limited model, the radio free-free emission from $\sim1$ to $300\:$GHz from the shock-ionized gas is in the range of $\sim 0.01$ to 1~mJy, i.e., generally increasing with evolutionary stage from 1.4 to $24\:M_\odot$, and increasing with the scale of the region considered from 1,000 to 25,000~au away from the protostar. At the inner scale of $\sim 1,000\:$au, these models become optically thick at frequencies below $\sim 0.1\:$GHz, while on the largest scale, they remain mostly optically thin, i.e., with flat radio spectral indices. Models without the cooling limit have much more ionized gas and radio luminosities that are greater by factors of $\sim 10$ to 1000, depending on the scale.

    \item In all cases the emission from shock ionized gas is expected to dominate over that from photoionization for protostellar luminosities $\lesssim 10^4\:L_\odot$. However, the precise transition from shock to photo-ionization dominance depends on details of the shock ionization model, i.e., treatment of cooling, and the protostellar formation conditions, i.e., accretion history as influenced by clump mass surface density.

    \item A comparison of our models with observed protostars shows that our fiducial cooling limited case has radio luminosities similar to that seen in low and intermediate mass sources by \citet{Anglada1995}, but about 10 to 100 times lower than the massive protostars studied in the SOMA Survey \citep{Roseroetal2019}. Possible explanations for this include a more limited role for cooling in limiting the extent of post-shock ionization and/or a significant contribution from photoionization. More detailed comparisons, e.g., of full radio spectra and spectral index maps of jet structures, are needed to distinguish among these possibilities.

    \item Our models produce 10-year variability in 5.3~GHz emission at the level of up to $\sim5\%$ on scales $\sim 1,000\:$au to 25,000~au. Such short timescale variability is associated with changing gas conditions in the fast moving outflow. Similar levels of variability have been seen to be typical in a sample of HC HII regions in W49A by \citet{dePreeetal2018}, which is strong evidence in favor of their ionized gas being associated with protostellar outflows. The higher level of variability, up to $\sim40\%$, in a couple of sources may indicate a role for short timescale accretion/outflow variability, which has not yet been included in our modeling.

\end{enumerate}

\section*{Acknowledgements}
The authors thank Rub\'{e}n Fedriani for discussions regarding his ionization fraction observations and Viviana Rosero for discussions regarding the SOMA-Radio observations. This research was initiated in the Virginia Initiative on Cosmic Origins (VICO) summer undergraduate student research program
at the University of Virginia (UVA), Charlottesville, VA, USA, including a collaborative visit to the Chalmers Initiative on Cosmic Origins (CICO), Chalmers University of Technology, Gothenburg, Sweden. This work was also supported by the National Science Foundation (NSF) collaborative grant ``Protostellar Jets Across the Mass Spectrum'' (AST-1910675).
The authors acknowledge the use of the Rivanna computing cluster, under the management of Research Computing at UVA.
JCT acknowledges support from NSF grants AST-1910675 and AST-2206450 and ERC Advanced Grant 788829 (MSTAR). JES acknowledges support from NSF grant AST-1910675.
JPR acknowledges support from VICO, the NSF under grant nos.\ AST-1910106 and AST-1910675, and NASA via the Astrophysics Theory Program under grant no.\ 80NSSC20K0533.

\software{ 
jupyter \citep{Kluyveretal2016_jupyter}, 
matplotlib \citep{Hunter2007_matplotlib},
numpy \citep{numpy2011},
pillow \citep{clark2015pillow}
}

\newpage
\appendix
\renewcommand{\thetable}{\Alph{section}\arabic{table}}
\renewcommand{\thefigure}{\Alph{section}\arabic{figure}}    
\setcounter{table}{0}
\setcounter{figure}{0}


\section{Calculation of Shock Ionization}
\label{app:method}

We refer to coordinates $(z,x,y)$ indexed by $(i,j,k)$ where $z$ is the outflow direction. The methods are general for emission in any principle axial direction, but we focus specifically on adopting the $y$-direction as our line of sight direction and map $x$-$z$ plane projections.

Shock velocities are calculated by taking, at each face of a cell, the velocity difference between the cell of interest and its neighbors, modulo only using converging velocity components. This yields the inward velocity to the current cell, in the reference frame of the gas in the cell. If this value is net positive in the inward direction (i.e., converging), it is set as the shock velocity at that face, otherwise, we take a shock velocity of zero. While these velocity differences will only be true shock velocities if they are larger than the sound speed, applying the shock calculation also to values below the sound speed has a negligible effect on the shock temperature. This process was replicated for every cell of the MHD simulation snapshot. 

Given the shock velocities, a post-shock temperature is calculated for the shock at each face, according to the standard Rankine-Hugoniot shock jump conditions for the case that the Mach number, ${\cal M}$, is $\gg 1$: 
\begin{eqnarray}
     T &=& \frac{2(\gamma-1)}{(\gamma +1)^2} \frac{\mu v_s^2}{k_B} = 
     \frac{3}{16} \frac{\mu v_s^2}{k} \nonumber \\
     &=& 1.38 \times 10^7 \left(\frac{\mu/m_{\rm H}}{1.4/2.3}\right)\left(\frac{v_s}{1000 \mathrm{km/s}}\right)^2 \mathrm{K},
   \label{eq:shock_temp}
\end{eqnarray}
where $\gamma = \frac{5}{3}$ is the ratio of specific heats, $\mu$ is the mass per particle and $v_s$ is the shock velocity.

The ionization fraction at each face from collisional ionization is calculated via \citep[e.g.,][]{Draine2011}:
\begin{equation}
    \frac{n(\mathrm{H})}{n(\mathrm{H^+)}} = \frac{\langle \sigma v\rangle_{\rm rr}}{\langle\sigma v\rangle_{\rm ci}} = \frac{2^4}{3^{3/2}}\left(\frac{e^2}{\hbar c}\right)^3 \frac{B}{k_BT} e^{B/kT};
  \label{eq:neutralfrac}
\end{equation}
\begin{equation}
    \chi_\mathrm{\rm{H}+} = \frac{n(\rm{H}^+)}{n(\rm{H}) + n(\rm{H}^+)} = \left(1+ \frac{n(\rm{H})}{n(\rm{H}^+)} \right)^{-1}
  \label{eq:ionizationfrac_H+}
\end{equation}
where  $B = (157,800 \mathrm{K}) k$ is the ionization energy for Hydrogen, $n(\rm{H})$ is the number density of hydrogen, $n(\rm{H^+})$ is the number density of hydrogen ions, $\langle \sigma v\rangle_{\rm rr} $ and $\langle \sigma v\rangle_{\rm ci}$ are the radiative recombination and collisional ionization rates, respectively.
A temperature floor of 300 K for ionization is set, such that for cell temperatures below this value the ionization is taken to be zero since it can conservatively be assumed that any cell with a temperature below this value has zero ionization. 

With temperatures and ionization fractions calculated for the converging flow of each cell face, a single average value is determined for each cell by weighting each face's 
contribution by the associated mass flux.

\subsection{Cooling Effects}
\label{sec:meth_cooling}
The shocks in our simulation are modeled adiabatically, i.e., using an adiabatic equation of state to relate the pressure and temperature. For our initial approximation (Case A), the resulting temperatures and ionization fractions are assumed to fill the entire cell. This approximation is reasonable as long as the shocked gas fills the cell before significant cooling takes place. 

However, for Case B, which is our fiducial case, we consider the effects of cooling.
The cooling times can be calculated via \citep[e.g., eq. 35.34 of][]{Draine2011}:
\begin{equation}
    t_\mathrm{cool} = \frac{3 n_\mathrm{H}  k T}{2 \Lambda(T)}
    \label{eq:t_cool}
\end{equation}
where the radiative cooling function at low temperatures ($T<10^5K$) is approximated as
\begin{equation}
    \Lambda(T) \approx (3.98\times 10^{-30} \mathrm{erg\ cm^3\ s^{-1}}) (T/\mathrm{K})^{1.6} n_\mathrm{H} n_\mathrm{e},
    \label{eq:Lambda(lowT)}
\end{equation}
and at high temperatures ($T > 10^5\mathrm{K}$) as:
\begin{equation}
    \Lambda(T) \approx (1.1 \times 10^{-22}~ \mathrm{erg\ cm^3\ s^{-1}}) \Big( \frac{T}{10^6 \mathrm{K}}\Big)^{-0.7} n_\mathrm{H} n_\mathrm{e}.
    \label{eq:Lambda(highT)}
\end{equation}
Accuracy of cooling times is more important for the high-temperature regions, as these are where most free-free emission occurs.

The post-shock cooling times are then compared to the flow times, i.e., the time it takes the shocked gas to fill the cell, given as a mass flux weighted average of the converging shock's timescale from each face:
\begin{eqnarray}
    t_\mathrm{flow} &=& \frac{\sum_{v_{s}>0} |(\frac{1}{2}\Delta s) / (v_{s}/4)| v_\mathrm{n} \rho_\mathrm{n}}{v_\mathrm{n} \rho_\mathrm{n}} \nonumber \\
    &=& 2\frac{\sum_{v_{s}>0} |\Delta s / v_{s}| v_\mathrm{n} \rho_\mathrm{n}}{v_\mathrm{n} \rho_\mathrm{n}},
    \label{eq:t_flow}
\end{eqnarray}
where $v_s/4$ is the post-shock velocity and $\frac{1}{2}\Delta s$ is the distance from the face to the center of the cell. 

\begin{figure}
    \centering
    \includegraphics[width=\columnwidth]{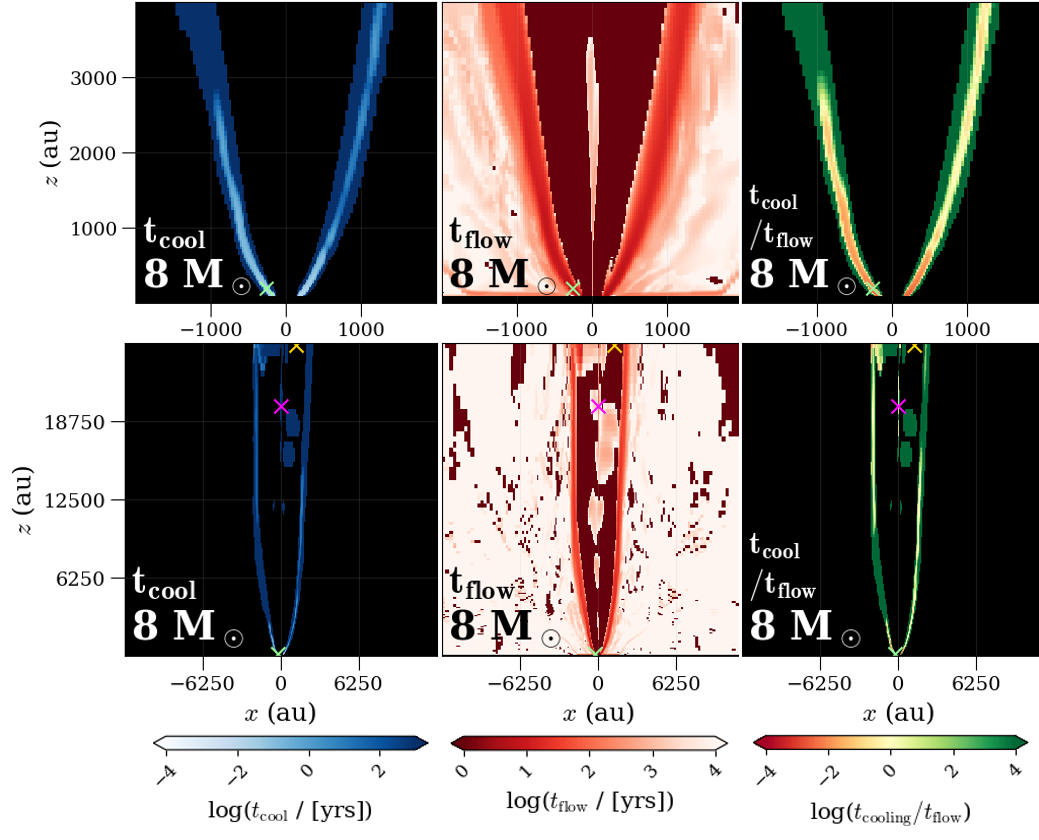} 
    \caption{
    Cooling time $t_\mathrm{cool}$ (\textit{left}), flow time $t_\mathrm{flow}$ (\textit{center}), and their ratio $t_\mathrm{cool}/t_\mathrm{flow}$ (\textit{right}) for a central slice of the 39,000 yr, 8~$M_\odot$ snapshot are presented for the region $-2000<x<2000$, $0<z<4000$ in the top row and for $-12500<x<12500$, $0<z<25000$ in the bottom row. Diverging color scales in the ratio maps show most of the region to have larger cooling times, with the exception of several strongly-emitting regions at the edges of the jet. The $\times$ symbols mark several test cells discussed in \S\ref{sec:app_example}.}
    \label{fig:cooling}
\end{figure}

Maps of the cooling time $t_\mathrm{cool}$, the flow time $t_\mathrm{flow}$, and their ratio $t_\mathrm{cool}/t_\mathrm{flow}$ are shown in Fig.\ \ref{fig:cooling} for a slice from the snapshot corresponding to $m_*=8\:M_\odot$. It shows that there are significant regions with cooling times 10 times shorter than the corresponding flow times, thus motivating Case B in preference to Case A.

For Case B we modify the 1-D radiative transfer integration calculation in the following way. Every cell along the column of integration with a cooling time less than its flow time, has its integration depth $\Delta y$ scaled by that cell's ratio of $t_\mathrm{cool}/t_\mathrm{flow}$, as explained in Section \ref{sec:meth_freefree}. In effect, this calculates emission only from the volume of the cell flooded by shocked gas before significant cooling has taken place. Results comparing Cases A and B are presented in Appendix \ref{app:caseA}. 

\subsection{Calculating Free-Free Radio Emission}
\label{sec:meth_freefree}
With the aforementioned mass flux weighted temperatures $T$ and ionization fractions $\chi_\mathrm{H+}$, we then calculate the free-free radio emission due to shocks in the simulated outflow. The free-free emission coefficients $j_\nu$, absorption coefficients $\kappa_\nu$, and optical depths in each axial direction $\tau_{\nu,z}, \tau_{\nu,x}, \tau_{\nu,y}$ are \citep[e.g.,][]{Draine2011}:
\begin{eqnarray}
    j_\nu = 3.86 g_\mathrm{ff} \frac{e^6}{m_e^2c^3}\left(\frac{m_e}{kT}\right)^{1/2} n_e n_p e^{-h\nu/(kT)} \nonumber \\
    \mathrm{[erg \ cm^{-3} \ sr^{-1} \ \mathrm{Hz}^{-1} \ \mathrm{s}^{-1}]};
    \label{eq:j_nu}
\end{eqnarray}
\begin{equation}
    \kappa_\nu = \frac{4}{3}\left( \frac{2\pi}{3}\right)^{1/2} \frac{n_en_pe^6g_\mathrm{ff}}{m_e^{3/2}c(kT)^{3/2}}\frac{1}{\nu^2}\ [\mathrm{cm}^{-1}]
    \label{eq:kappa_nu};
\end{equation}
\begin{equation}
    \tau_{\nu,z} = \kappa_\nu \Delta z; \quad \tau_{\nu,x} = \kappa_\nu \Delta x; \quad \tau_{\nu,y} = \kappa_\nu \Delta y,
    \label{eq:tau}
\end{equation}
where $g_\mathrm{ff} = 5.96 (T/10^4 \mathrm{K})^{0.15} (\nu / \mathrm{GHz})^{-0.1} $ is the Gaunt factor and we assume $n_e = n_p = n_\mathrm{H+} = (\chi_\mathrm{H+} \rho)/\mu_{\rm H}$, where $\mu_{\rm H} = 1.4 m_{\rm H} = 2.34 \times 10^{-24}\:$g is the mass per H nucleus, i.e., assuming $n_{\rm He}=0.1 n_{\rm H}$. Note, our analysis ignores ionization of He, i.e., it assumes He is not ionized to a significant level.

For Case A, after determining the local emission and absorption coefficients for each cell, and assuming a uniform temperature within each cell, we calculate specific intensity, $I_\nu$, via integrating the 1D radiative transfer equation:
\begin{eqnarray}
    I_\nu(s) &=& I_\nu(0)e^{-\tau_\nu} + \int_{0}^{s} ds' j_\mathrm{ff} e^{-[\tau_\nu(s) - \tau_\nu(s')]}  \nonumber \\ 
    &=& I_\nu(0)e^{-\tau_\nu} + \int_0^{\tau_\nu} \left[\frac{j_{\mathrm{ff},\nu}}{\kappa_{\mathrm{ff}, \nu}}\right] e^{-(\tau-\tau')}d\tau'.
    \label{eq:1Dradtransfer}
\end{eqnarray}
In this work, we exclusively adopt the $y$-direction in the simulation snapshots as the line of sight. 
This analysis then yields maps of the emission from the simulation domain in the $x$-$z$ plane. 


For Case B, i.e., accounting for cooling effects, the radiative transfer equation is modified
by scaling the integration depth through each cell
by the ratio $t_\mathrm{cool}/t_\mathrm{flow}$ if this ratio is less than 1. 
This serves to include emission only from the portion of each cell flooded by shocked gas before significant cooling takes place.

Integrated fluxes were calculated for different-sized regions prescribed by $-r/2 < x < r/2$ and $0<z<r$, for $r$ = 1000, 2000, 4000, 8000, and 16000 au, as well the full simulation region, which we refer to as 25000 au because it extends to 25000 au in the $z$ direction. Although, the entire simulation region extends past $25000 / 2$ au to $\pm$15000 au in the $x$ and $y$ directions, by this point the $x$ and $y$ regions are large enough to cover the entire jet regardless, so it is the $z$ range that is of most importance in characterizing how much of the jet is included. These fluxes were calculated as a surface integral of the intensity over the solid angle region $\Omega = A/d^2$, 
\begin{equation}
    S_\nu \equiv \int I_\nu d\Omega = \sum_{i,j} I_{\nu,i,j} \frac{\Delta x_i \Delta x_j}{d^2},
\end{equation}
assuming a distance of $d$ = 1 kpc, then doubled to account for the opposite hemisphere of the expected bipolar outflow that is not included in the MHD simulation.
In doing so for each snapshot and multiple frequencies, we obtain spectra and fluxes for a given frequency at a given time of the simulation.


\subsection{Single Cell Examples}
\label{sec:app_example}

To demonstrate our methods on a simulation cell-by-cell basis, we consider a partially ionized cell from the 39,000-year ($8\:M_\odot$) snapshot. This particular cell has indices $(i,j,k) = (156, 141, 139)$ in the 3D snapshot, and is marked with a fuchsia $\times$ in Fig.\ \ref{fig:cooling}. The relevant simulation data for cell $(156,141,139)$ and each of its neighboring cells are given in columns 1 through 9 of the top section of Table \ref{tab:test_cell}. Only the neighboring cells' velocities that flow in the direction to/from the test cell are listed, as the other velocities have no impact on the test cell of interest. The velocity difference, shock velocity, shock temperature, and ionization fraction for each face are then given in columns 10-13 of Table \ref{tab:test_cell}, calculated according to Eqs.\ (\ref{eq:shock_temp})-(\ref{eq:ionizationfrac_H+}). Averaging the face temperatures $T_f$ and face ionization fractions $\chi_{\mathrm{H+},f}$ by the mass-fluxes $\rho v_f$
yields flux-weighted average values of $T=19,300\ \mathrm{K}$ and $\chi_\mathrm{H+}=0.60$ for the cell. 

Since this temperature is less than $10^5\:$K, we use Eqs.\ (\ref{eq:Lambda(lowT)}) and (\ref{eq:t_cool}) with $n_\mathrm{H} = \rho/\mu_\mathrm{H}$ to find a cooling time of $t_\mathrm{cool} = 1025\ \mathrm{yr}$. For the flow time, we average the timescales from all sides with non-zero shock velocities. In this example, those are cells $i-1$, $i+1$, and $j-1$. This cell has $\Delta z =7.24\times10^{10}$ km and  $\Delta x = \Delta y =  1.84\times10^{9}$ km. Eq.\ \ref{eq:t_flow} gives time scales of 210 yr, 137 yr, and 614 yr, respectively. The flux-weighted average of these values yields $t_\mathrm{flow} = 169\ \mathrm{yr}$. Thus, we find a ratio $t_\mathrm{cool}/t_\mathrm{flow} = 1025/169 \sim 6.1$, indicating that the cooling time is $\sim$ 6.1 times longer than the flow time, so it is reasonable in this case to assume the shock variables flood the cell before significant cooling occurs. 

Cell $(i,j,k) = (7,122,139)$ serves as a counter-example, where the flow timescale exceeds the cooling timescale. The cell is marked with a green $\times$ in Fig. \ref{fig:cooling} and the relevant data for each face are found in the middle section of Table \ref{tab:test_cell}.  Averaging the $T_f$ and $\chi_{\mathrm{H+},f}$ values by the mass flux 
yields $T=29,300\ \mathrm{K}$ and $\chi_\mathrm{H+}=0.999$ for the cell. Since this temperature is less than $10^5$K, we use Eqs.\ (\ref{eq:Lambda(lowT)}) and (\ref{eq:t_cool}) to find $t_\mathrm{cool} = 0.024\ \mathrm{yr}$. To calculate the flow time, we consider the sides with non-zero shock velocity, i.e.,  $j-1$, and $j+1$, with cell width $\Delta s = \Delta x =2.77\times10^{9}$ km. Eq.\ (\ref{eq:t_flow} gives time scales of 4.88 yr and 4.87 yr, respectively, with a mass flux weighted average of $t_\mathrm{flow} \sim 4.9~\mathrm{yr}$. The ratio of $t_\mathrm{cool}/t_\mathrm{flow} = 0.005$ indicates that significant cooling will occur before the shocked gas floods the cell.

Finally, we consider an especially hot, mostly (99\%) ionized test cell with indices $(i,j,k) = (165,191,139)$. The cell is marked in yellow in Fig. \ref{fig:cooling} and the relevant data for each face are found in the lower section of Table \ref{tab:test_cell}.  Averaging the $T_f$ and $\chi_{\mathrm{H+},f}$ values by the mass flux yields $T=4,565,000\ \mathrm{K}$ and $\chi_\mathrm{H+}=0.99$ for the cell. Since this temperature is greater than $10^5$K, we use Eqs.\ (\ref{eq:t_cool}) and (\ref{eq:Lambda(highT)}) to find the cooling timescale, $t_\mathrm{cool} = 4676\  \mathrm{yr}$. The sides with non-zero shock velocity are now only $i-1$, and $i+1$, with cell height $\Delta s = \Delta z =7.24\times10^{10}$ km. This results in flow time scales of 15.39 yr and 12.35 yr, respectively, and a flux-weighted average of $t_\mathrm{flow} = 12.75$ yr. Thus, we find a ratio $t_\mathrm{cool}/t_\mathrm{flow} = 4676/12.75 \sim 367$, indicating that the cooling time is $\sim$ 367 times the flow time, so it is again reasonable in this case to assume the shock floods the cell before significant cooling occurs. 

\begin{deluxetable*}{cccccccccccccc}
\tablecaption{Test Cell Simulation Data\label{tab:test_cell} } 
\tablehead{\colhead{$f$} & \colhead{($i,j,k)$} & 
\colhead{$z$} & \colhead{$x$} & \colhead{$y$} & \colhead{$\rho$} & \colhead{$v_z$} & \colhead{$v_x$} & \colhead{$v_y$} & \colhead{$v_f - v_\mathrm{cur}$} & \colhead{$v_s$} & \colhead{$T_f$} & \colhead{$\chi_{\mathrm{H+},f}$} & \colhead{$\rho v_f$}\\ 
 \colhead{(face)} & & \colhead{[au]} & \colhead{[au]} & \colhead{[au]} & \colhead{[g cm$^-3$]} & \colhead{[km/s]} & \colhead{[km/s]} & \colhead{[km/s]} & \colhead{[km/s]} & \colhead{[km/s]} & \colhead{[K]} & & \colhead{[g cm$^-3$ km/s]}
 }
\startdata
$(i,j,k)$	&	(156,141,139)	&	20,008	&	18	&	-6	&	1.68E-23	&	1316.82	&	2.86	&	-0.45	&		&		&		&		&		\\
$i-1$	&	(155,141,139)	&	19,530	&	18	&	-6	&	1.69E-23	&	1338.65	&		&		&	21.83	&	21.83	&	10,804	&	0.025	&	3.68E-22	\\
$i+1$	&	(157,141,139)	&	20,498	&	18	&	-6	&	1.62E-23	&	1283.43	&		&		&	-33.39	&	33.39	&	25,276	&	0.996	&	5.42E-22	\\
$j-1$	&	(156,140,139)	&	20,008	&	6	&	-6	&	1.20E-23	&		&	1.05	&		&	-1.81	&	0	&	0	&	0	&	0.00	\\
$j+1$	&	(156,142,139)	&	20,008	&	30	&	-6	&	2.18E-23	&		&	2.67	&		&	-0.19	&	0.19	&	1	&	0	&	4.14E-24	\\
$k-1$	&	(156,141,138)	&	20,008	&	18	&	-18	&	1.87E-23	&		&		&	-1.25	&	-0.8	&	0	&	0	&	0	&	0	\\
$k+1$	&	(156,141,140)	&	20,008	&	18	&	6	&	1.99E-23	&		&		&	0.32	&	0.77	&	0	&	0	&	0	&	0	\\
$(i,j,k)$	&	weighted	&	\nodata	&	\nodata	&	\nodata	&	\nodata	&	\nodata	&	\nodata	&	\nodata	&	\nodata	&	\nodata	&	19,332	&	0.601	&	\nodata	\\
\hline																											
$(i,j,k)$	&	(7,122,139)	&	195	&	-261	&	-6	&	3.39E-19	&	130.98	&	-96.36	&	-5.93	&		&		&		&	0	&		\\
$i-1$	&	(6,122,139)	&	181	&	-261	&	-6	&	4.31E-19	&	101.4	&		&		&	-29.58	&	0	&	0	&	0	&	0	\\
$i+1$	&	(8,122,139)	&	210	&	-261	&	-6	&	2.79E-19	&	161.48	&		&		&	30.50	&	0	&	0	&	0	&	0	\\
$j-1$	&	(7,121,139)	&	195	&	-280	&	-6	&	4.77E-19	&		&	-60.44	&		&	35.92	&	35.92	&	29,252	&	0.999	&	1.71E-17	\\
$j+1$	&	(7,123,139)	&	195	&	-243	&	-6	&	2.75E-19	&		&	-132.35	&		&	-35.99	&	35.99	&	29,366	&	0.999	&	9.90E-18	\\
$k-1$	&	(7,122,138)	&	195	&	-261	&	-18	&	3.22E-19	&		&		&	-11.79	&	-5.86	&	0	&	0	&	0	&	0	\\
$k+1$	&	(7,122,140)	&	195	&	-261	&	6	&	3.75E-19	&		&		&	0.51	&	6.44	&	0	&	0	&	0	&	0	\\
$(i,j,k)$	&	weighted	&	\nodata	&	\nodata	&	\nodata	&	\nodata	&	\nodata	&	\nodata	&	\nodata	&	\nodata	&	\nodata	&	29,294	&	0.999	&	\nodata	\\
\hline																											
$(i,j,k)$	&	(165,191,139)	&	24,860	&	1,261	&	-6	&	3.94E-22	&	821.22	&	66.03	&	-0.03	&		&		&		&		&		\\
$i-1$	&	(164,191,139)	&	24,268	&	1,261	&	-6	&	3.62E-22	&	1190.02	&		&		&	368.80	&	368.8	&	3,083,619	&	1.000	&	1.34E-19	\\
$i+1$	&	(166,191,139)	&	25,465	&	1,261	&	-6	&	1.90E-21	&	361.52	&		&		&	-459.70	&	459.7	&	4,791,019	&	1.000	&	8.75E-19	\\
$j-1$	&	(165,190,139)	&	24,860	&	1,217	&	-6	&	3.11E-22	&		&	64.51	&		&	-1.52	&	0	&	0	&	0	&	0	\\
$j+1$	&	(165,192,139)	&	24,860	&	1,305	&	-6	&	5.85E-22	&		&	67.01	&		&	0.98	&	0	&	0	&	0	&	0	\\
$k-1$	&	(165,191,138)	&	24,860	&	1,261	&	-18	&	3.89E-22	&		&		&	-0.93	&	-0.90	&	0	&	0	&	0	&	0	\\
$k+1$	&	(165,191,140) 	&	24,860	&	1,261	&	6	&	4.16E-22	&		&		&	0.53	&	0.56	&	0	&	0	&	0	&	0	\\
$(i,j,k)$	&	weighted	&	\nodata	&	\nodata	&	\nodata	&	\nodata	&	\nodata	&	\nodata	&	\nodata	&	\nodata	&	\nodata	&	4,565,000	&	1.000	&	\nodata	\\
\enddata	
\tablecomments{Each test cell of interest $(i,j,k)$, is surrounded by neighboring cells $i-1$ and $i+1$ directly above and below it, respectively, in the $z$-direction, $j-1$ and $j+1$ left of and right of it in the $x$-direction, and $k-1$ and $k+1$ in front and behind it in the $y$-direction. For each neighboring cell, only data that contributes to the test cell of interest's flux-weighted temperature and ionization fraction is included, and all non-contributing entries are left blank. On the last line of each segment, the test cell's flux-weighted average temperature and ionization fractions are listed.
}
\end{deluxetable*}

\setcounter{table}{0}
\setcounter{figure}{0}
\section{Results for Case A (no cooling)}\label{app:caseA}

\begin{figure*}
    \centering
    \includegraphics[width=1.0\linewidth]{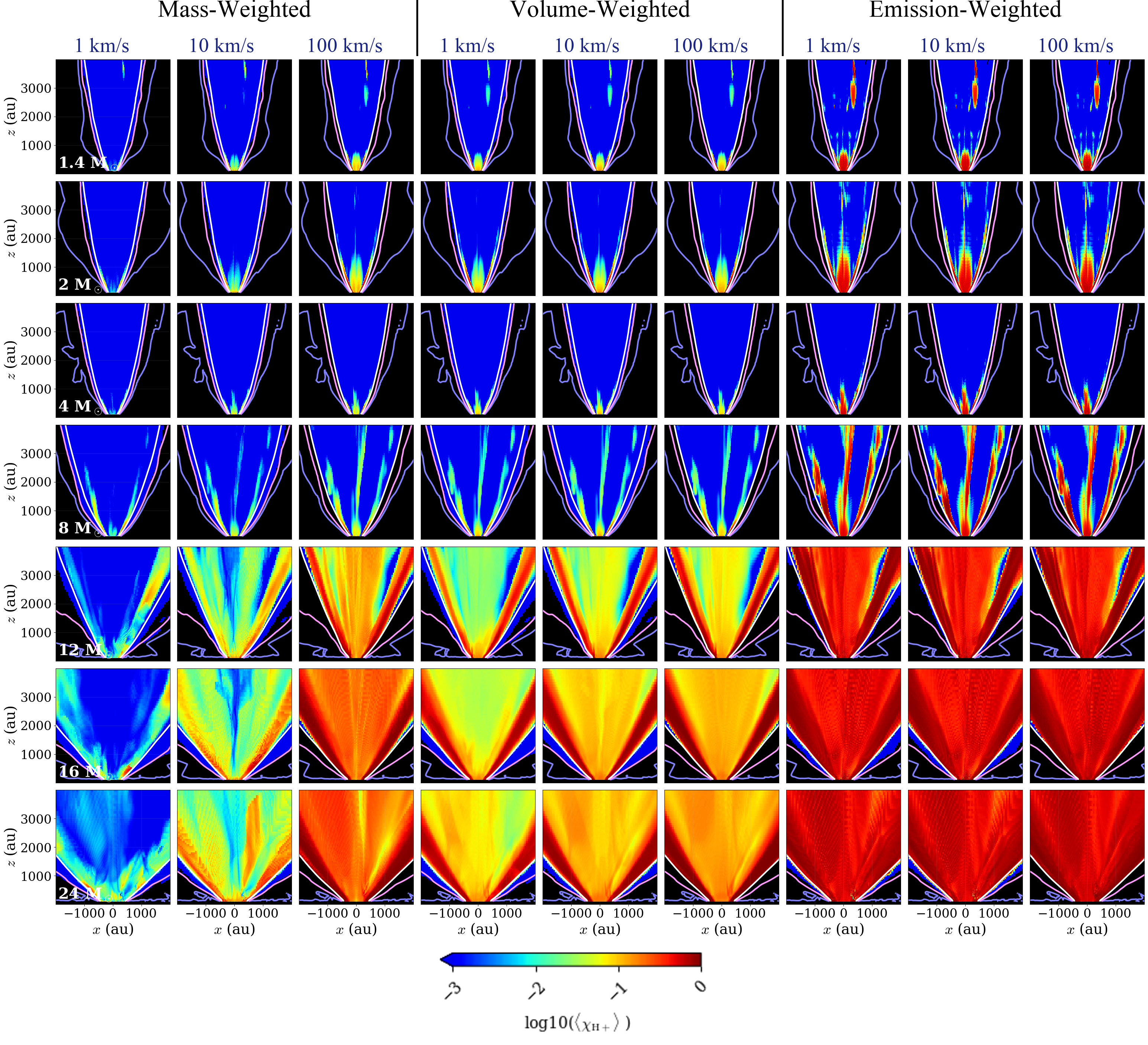}
    \caption{As Fig.~\ref{fig:chimaps_4000_ratio}, showing ionization fractions on the 4,000~au scale, but now for Case A (no cooling).
    \label{fig:chimaps_4000_noratio}}
\end{figure*}

\begin{figure*}
    \centering
    \includegraphics[width=1.0\linewidth]{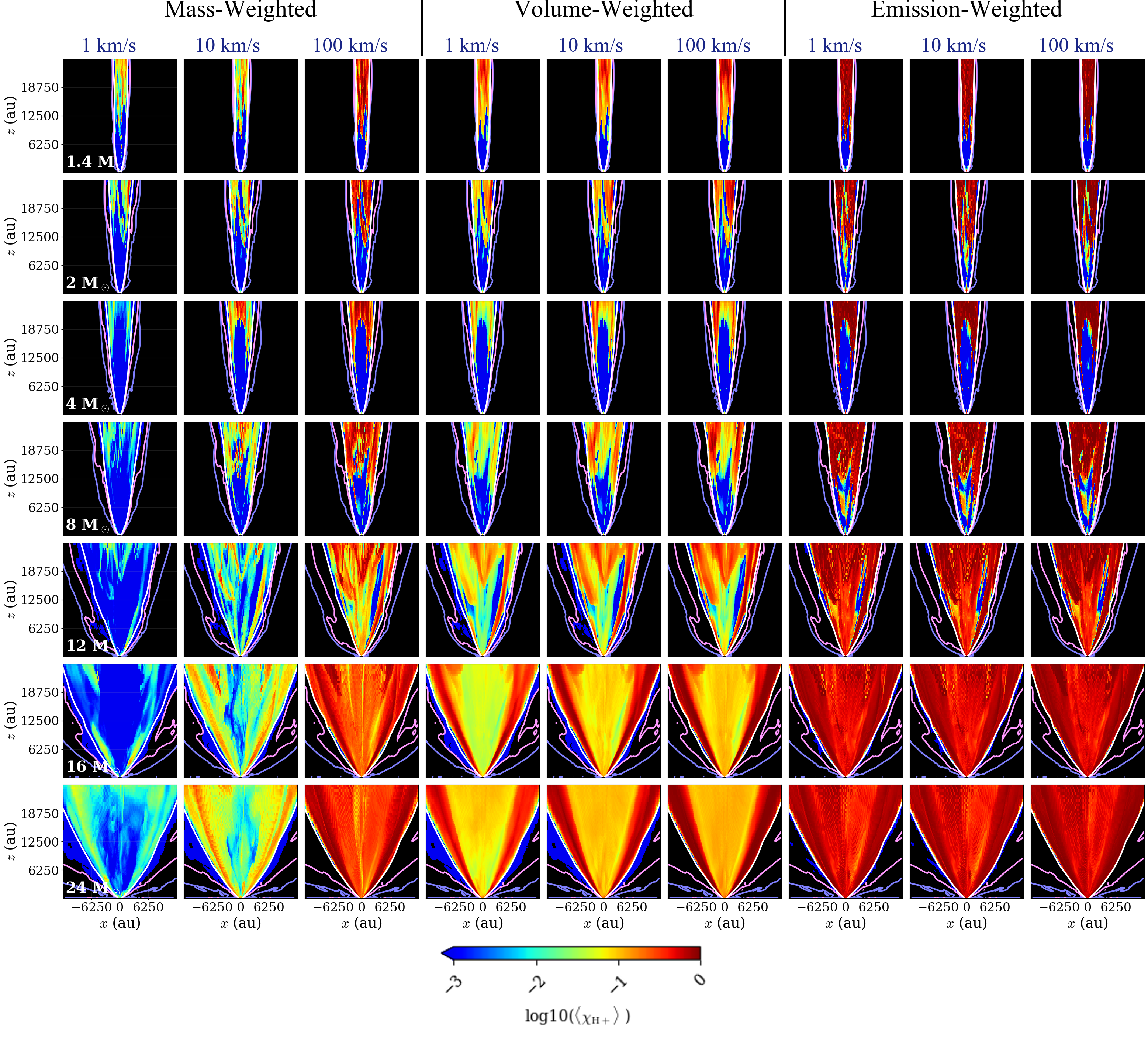}
    \caption{As Fig.~\ref{fig:chimaps_25000_ratio}, showing ionization fractions on the 25,000~au scale, but now for Case A (no cooling). 
    \label{fig:chimaps_25000_noratio}}
\end{figure*}

\begin{figure*}
\includegraphics[width=1.0\textwidth]{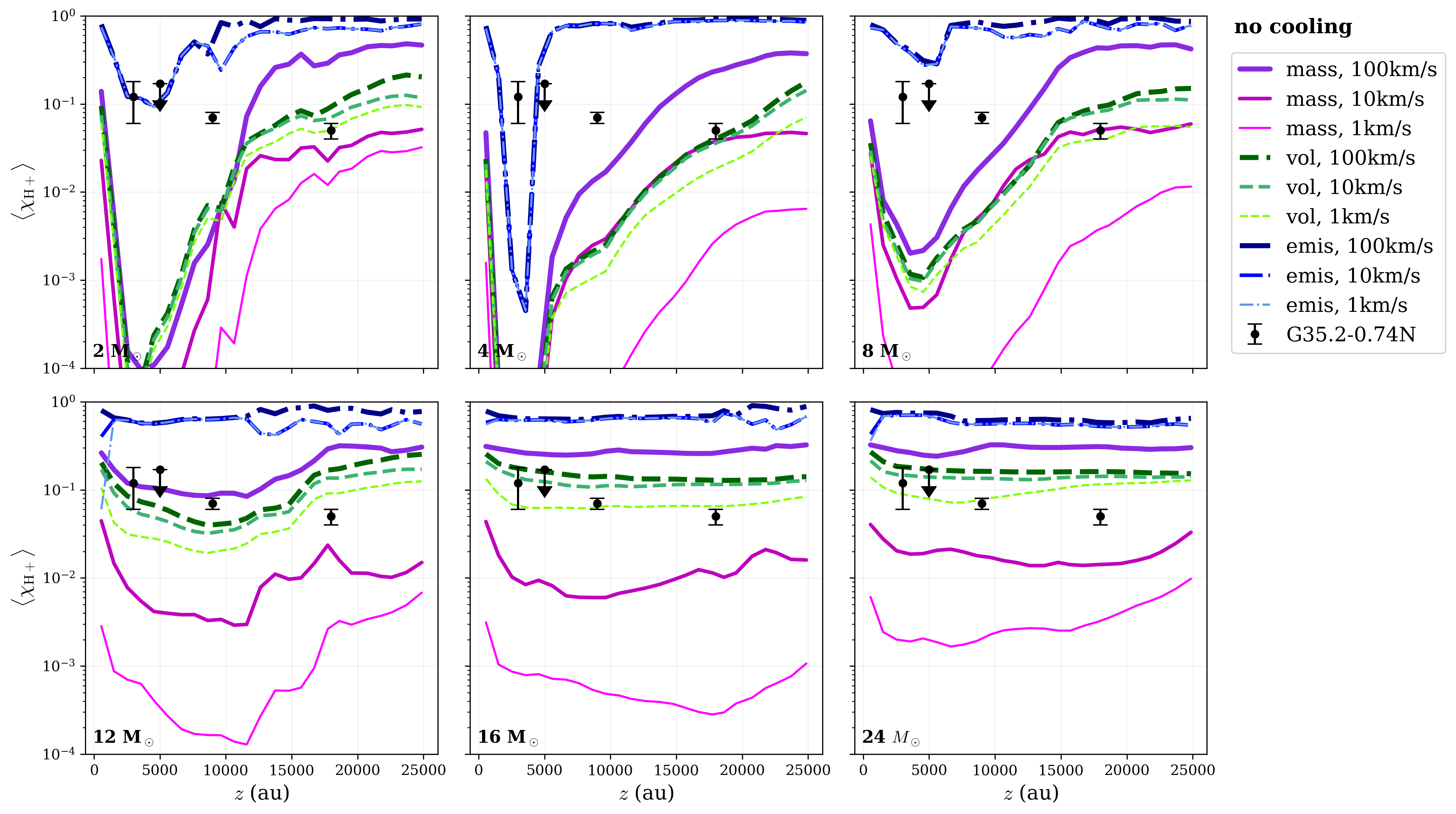}
\caption{As Fig.~\ref{fig:chiprofiles_ratio}, but now for Case A (no cooling).
\label{fig:chiprofiles_noratio}}
\end{figure*}

\begin{figure*}
    \centering
    \includegraphics[width=1.0\linewidth]{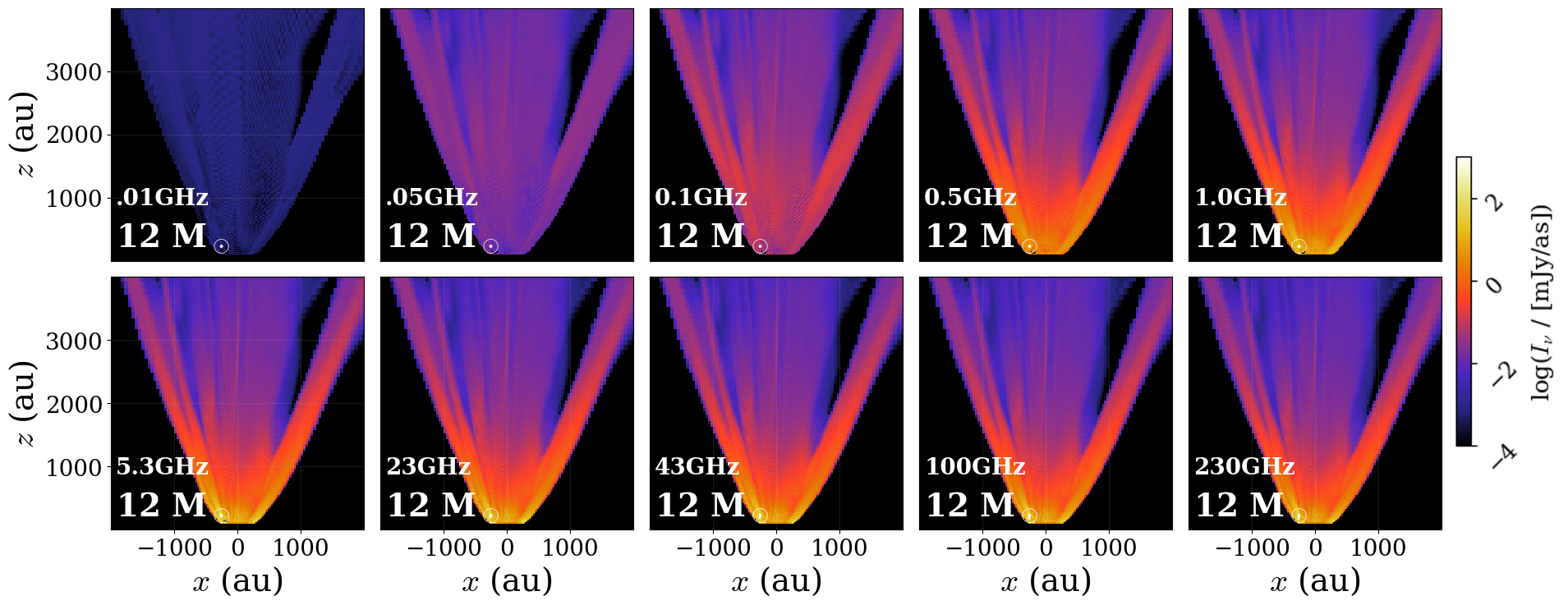}
    \includegraphics[width=1.0\linewidth]{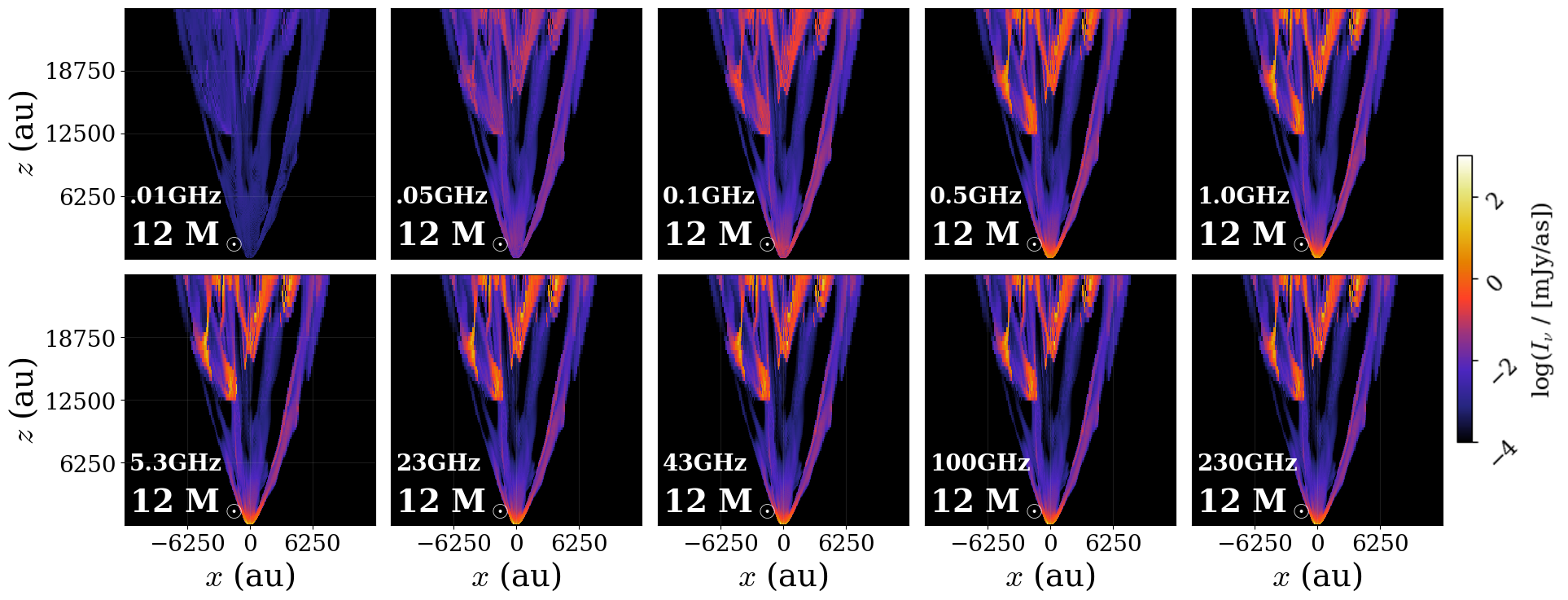}
    \caption{As Fig.~\ref{fig:intensitymaps_12M_ratio}, but now for Case A (no cooling).
    \label{fig:intensitymaps_12M_noratio}}
\end{figure*}

\begin{figure*}
    \centering
    \includegraphics[width=0.49\linewidth]{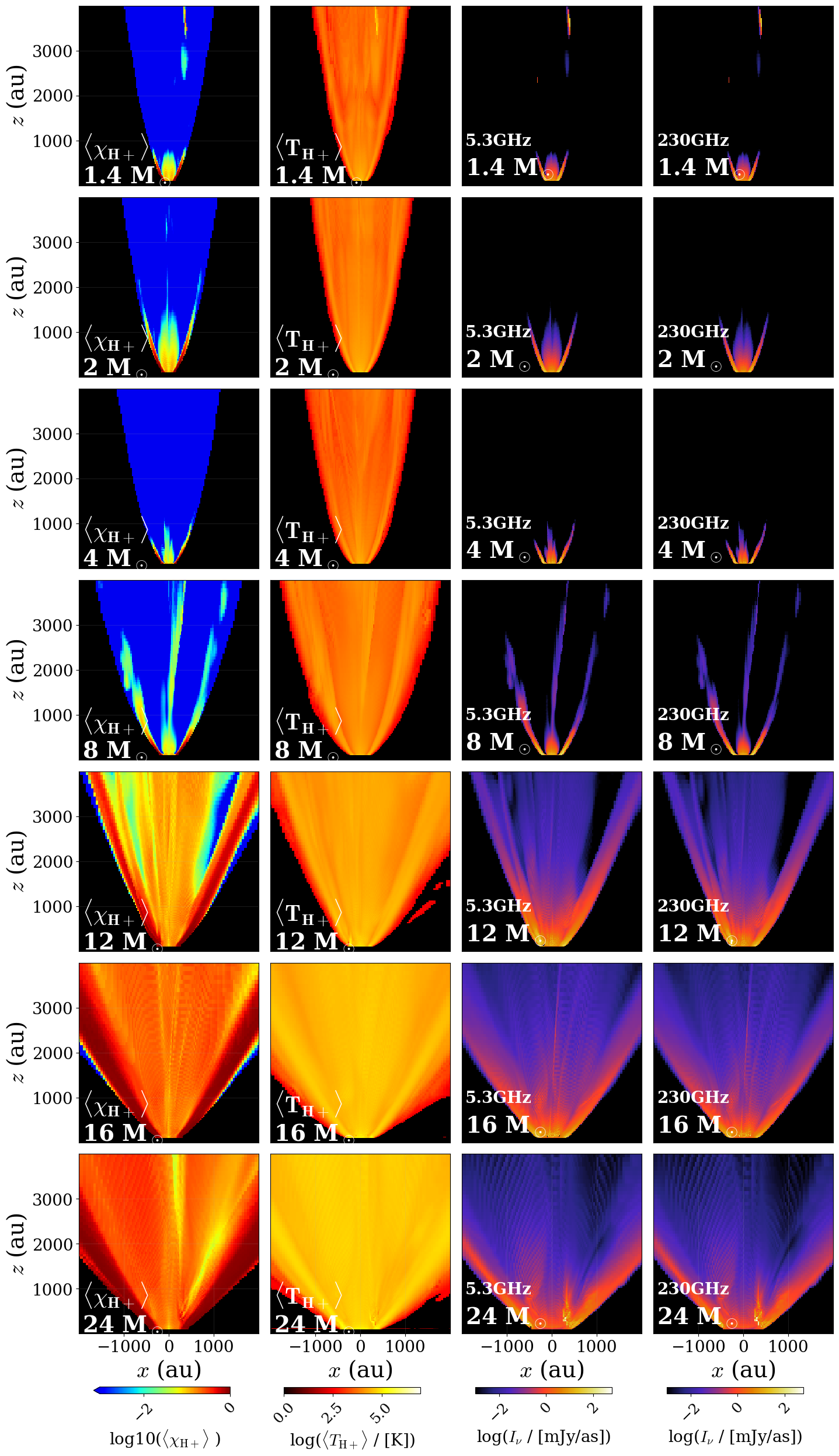}  
    \includegraphics[width=0.49\linewidth]{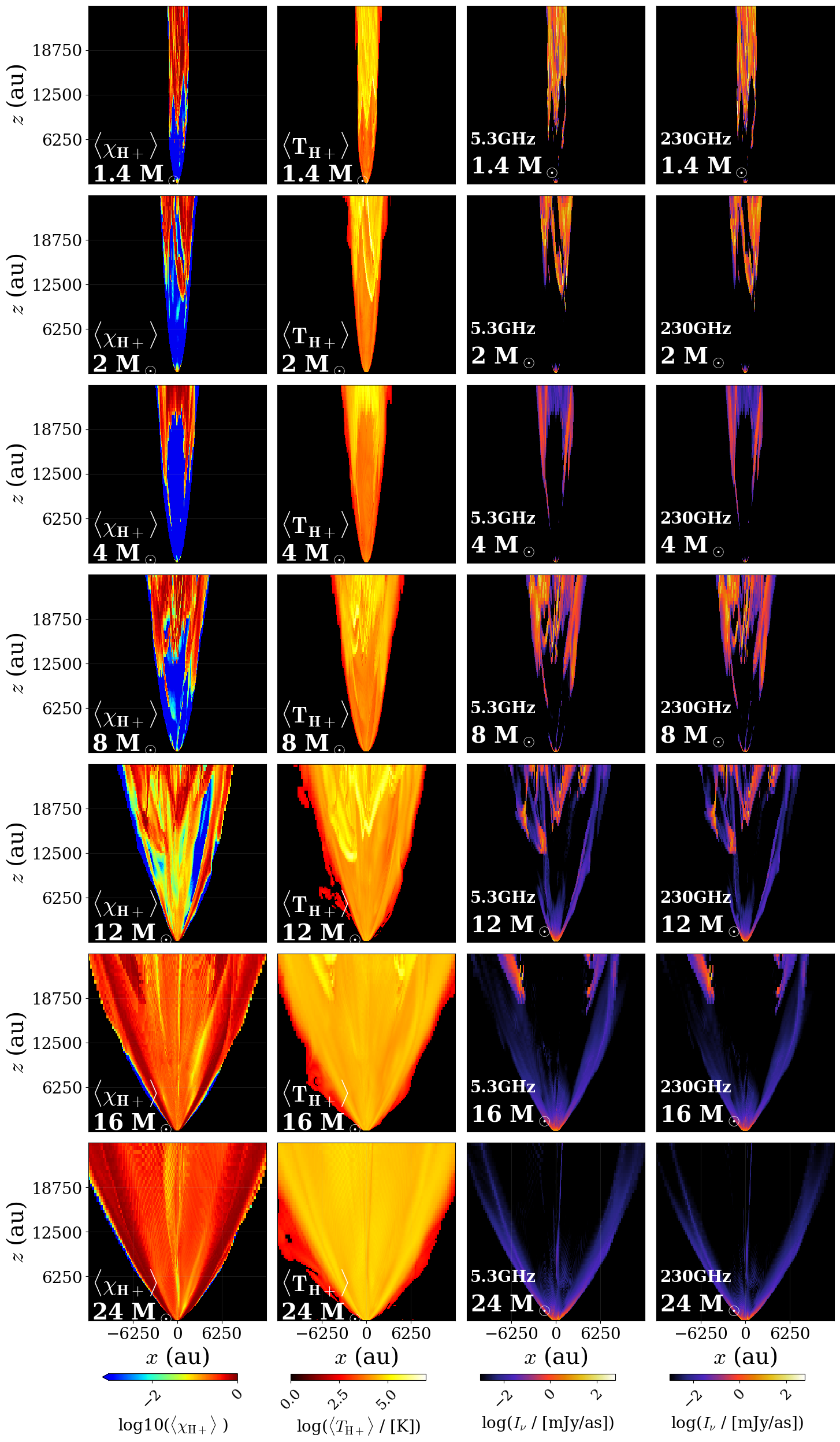}  
    \caption{As Fig.~\ref{fig:projections_ratio}, but now for Case A (no cooling).
\label{fig:projections_noratio}}
\end{figure*}

\begin{figure}
    \centering
    \includegraphics[width=\columnwidth]{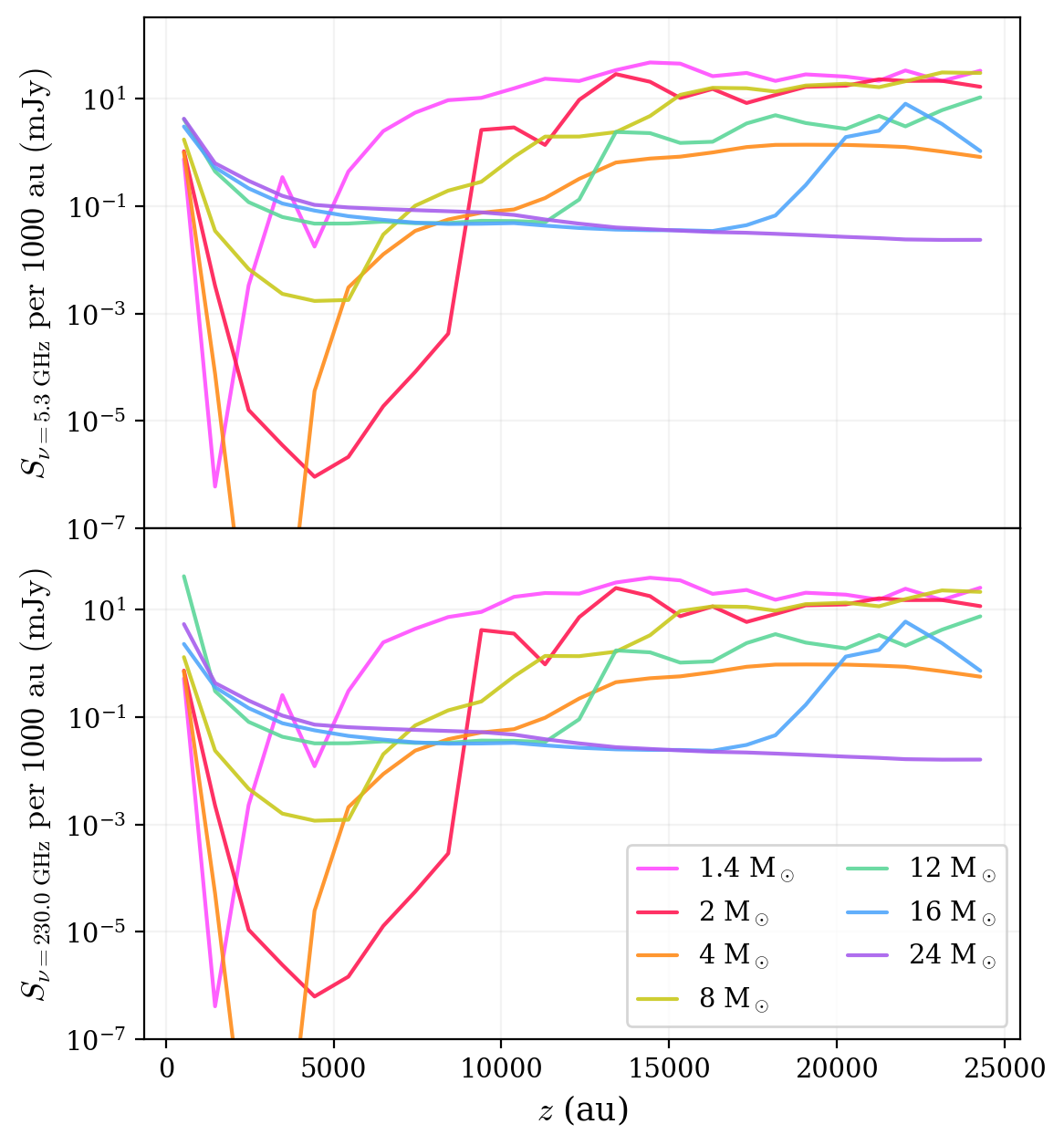}
    \caption{As Fig.~\ref{fig:fluxprofileB}, but now for Case A (no cooling).
    \label{fig:fluxprofileA}}
\end{figure}

\begin{figure*}
    \centering
    \includegraphics[width=\linewidth]{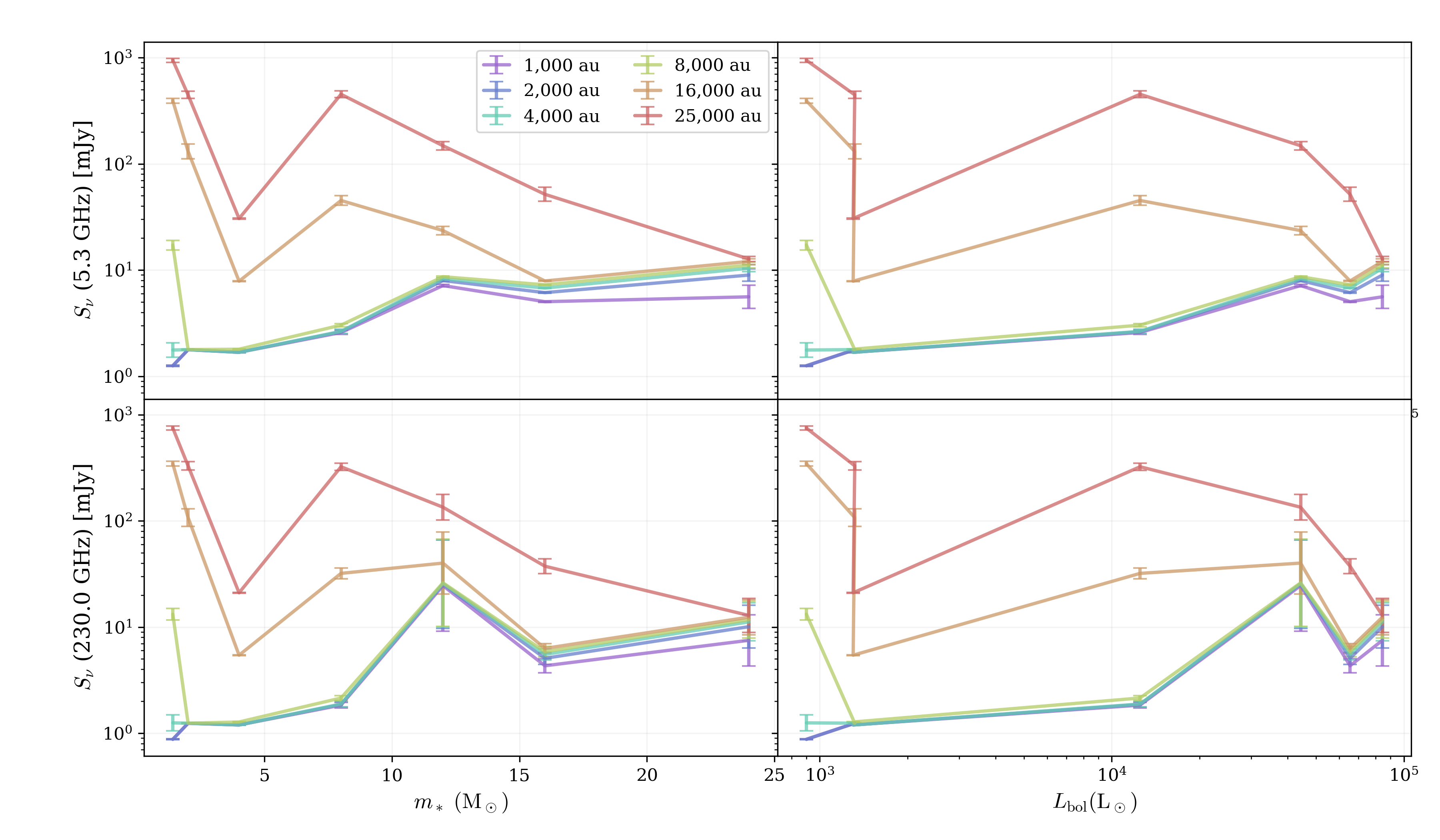}
    \caption{As Fig.~\ref{fig:fluxvsmass_caseB}, but now for Case A (no cooling).
\label{fig:fluxvsmass_caseA}}
\end{figure*}

\begin{figure*}
    \centering
    \includegraphics[width=1.0\textwidth]{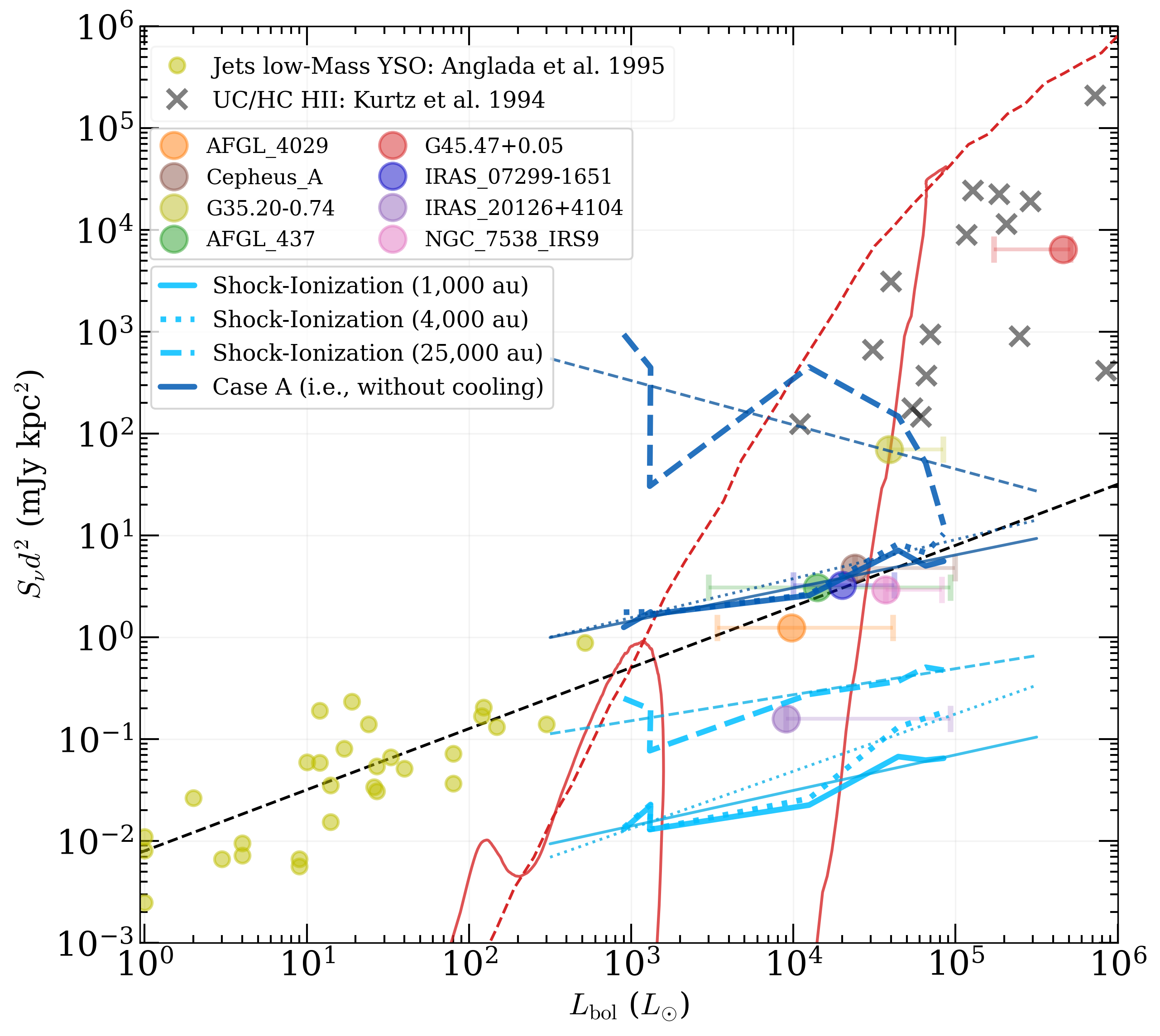}
    \caption{As Fig.~\ref{fig:AngladaDiagram}, but now also showing the results of Case A (no cooling) (dark blue solid, dotted and dashed lines for 1000~au, 4000~au, and 25000~au scales, respectively).
    \label{fig:AngladaDiagram_bothcases}}
\end{figure*}

\begin{figure}
    \centering
    \includegraphics[width=\columnwidth]{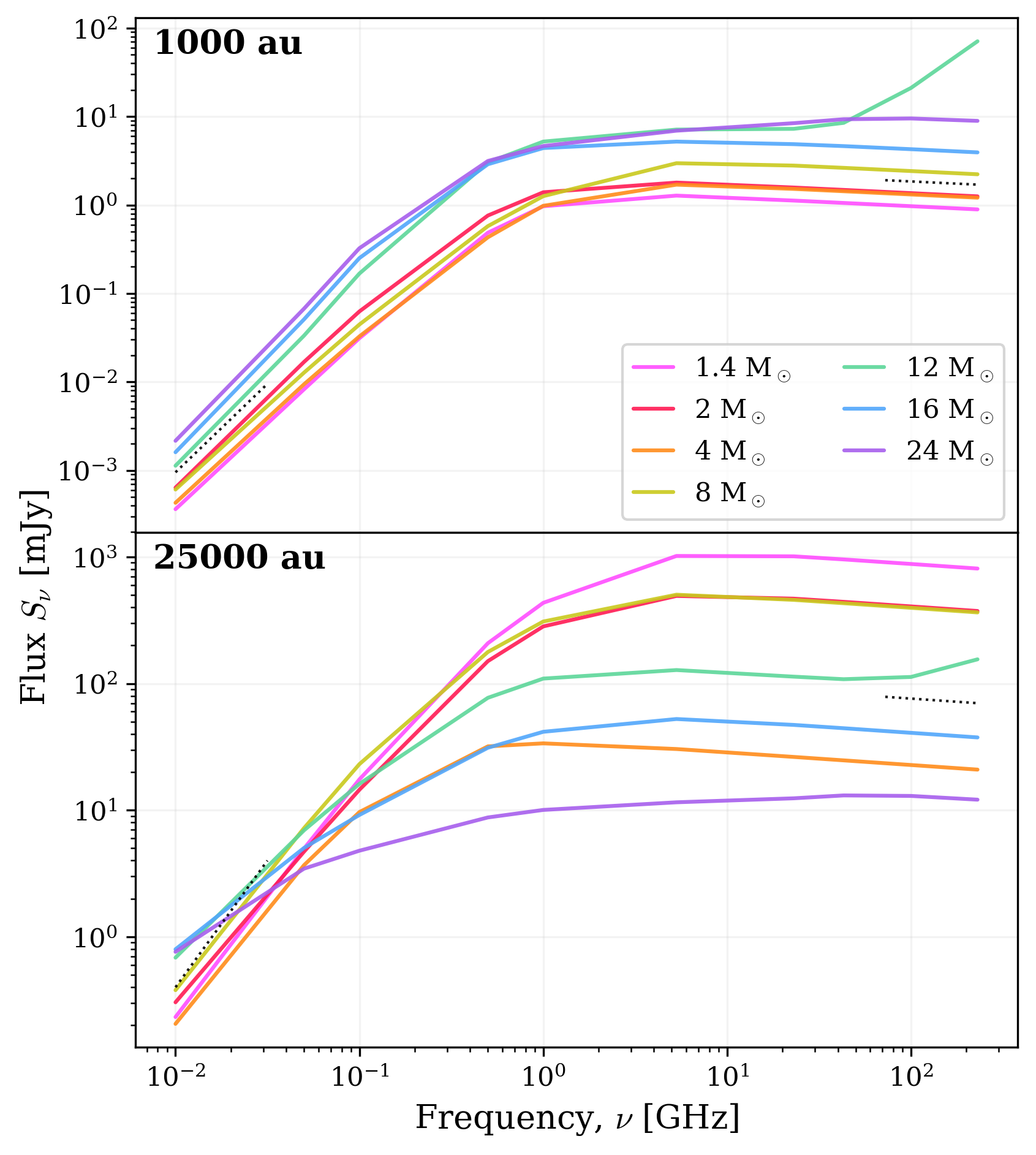}
    \caption{As Fig.~\ref{fig:spectralevolution_caseB}, but now for Case A (no cooling).
    }\label{fig:spectralevolution_caseA}
\end{figure}

\begin{figure}
    \centering
    \includegraphics[width=\columnwidth]{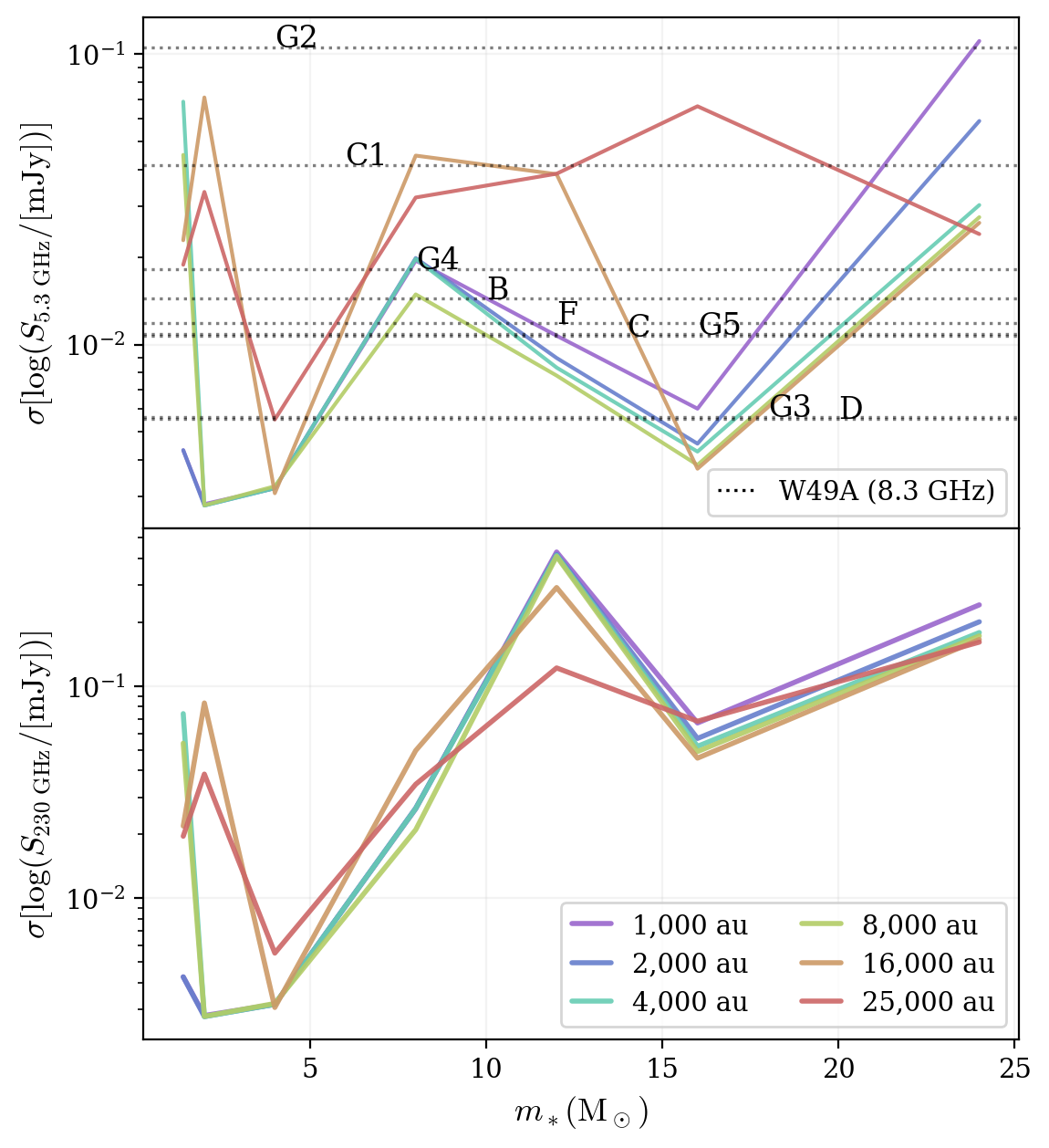}
    \caption{As Fig.~\ref{fig:percentvariation_caseB}, but now for Case A (no cooling).
    \label{fig:percentvariation_caseA}}
\end{figure}

In Figures~\ref{fig:chimaps_4000_noratio} to \ref{fig:percentvariation_caseA} we present results for Case A, i.e., without consideration of the cooling.

\begin{deluxetable*}{cc|ccc|ccc|ccc}
\tablecaption{Average Ionization Fractions for Case A, i.e., without cooling, corresponding to Figs.\ \ref{fig:chimaps_4000_noratio} and  \ref{fig:chimaps_25000_noratio}.\label{tab:avg_chi_noratio}} 
\tablehead{\colhead{Scale} & 
\colhead{$m_*$} & \multicolumn{3}{c}{Mass-Weighted} & \multicolumn{3}{c}{Volume-Weighted} & \multicolumn{3}{c}{Emission-Weighted} \\\colhead{[au]} &
 \colhead{[$M_\odot$]} & \colhead{$\geq$1 km/s} & \colhead{$\geq$10 km/s} & \colhead{$\geq$100 km/s} & \colhead{$\geq$1 km/s} & \colhead{$\geq$10 km/s} & \colhead{$\geq$100 km/s} & \colhead{$\geq$1 km/s} & \colhead{$\geq$10 km/s} & \colhead{$\geq$100 km/s}
 }
 
\startdata
	&	1.4	&	1.41e-04	&	4.07e-04	&	6.94e-01	&	8.23e-04	&	4.71e-04	&	6.94e-01	&	2.83e-03	&	6.49e-04	&	4.79e-01\\
	&	2.0	&	2.78e-04	&	8.52e-04	&	7.42e-01	&	1.07e-03	&	9.75e-04	&	7.42e-01	&	4.33e-03	&	1.29e-03	&	6.62e-01\\
	&	4.0	&	1.79e-04	&	2.13e-04	&	7.19e-01	&	5.17e-04	&	2.52e-04	&	7.19e-01	&	1.46e-03	&	3.71e-04	&	6.18e-01\\
4000	&	8.0	&	8.81e-04	&	6.45e-04	&	7.36e-01	&	1.41e-03	&	7.51e-04	&	7.36e-01	&	3.59e-03	&	1.07e-03	&	6.48e-01\\
	&	12.0	&	1.61e-03	&	1.25e-02	&	6.61e-01	&	8.06e-03	&	1.78e-02	&	6.64e-01	&	3.79e-02	&	2.41e-02	&	7.30e-01\\
	&	16.0	&	2.88e-03	&	3.85e-02	&	6.45e-01	&	1.50e-02	&	4.88e-02	&	6.50e-01	&	7.60e-02	&	5.75e-02	&	6.95e-01\\
	&	24.0	&	2.69e-03	&	4.12e-02	&	6.29e-01	&	1.78e-02	&	5.91e-02	&	6.42e-01	&	9.24e-02	&	7.53e-02	&	7.58e-01\\
\hline
	&	1.4	&	1.05e-03	&	1.06e-02	&	8.03e-01	&	2.37e-03	&	1.20e-02	&	8.03e-01	&	1.05e-02	&	1.89e-02	&	9.17e-01\\
	&	2.0	&	8.15e-04	&	7.85e-03	&	7.21e-01	&	1.80e-03	&	9.61e-03	&	7.21e-01	&	8.09e-03	&	1.44e-02	&	8.75e-01\\
	&	4.0	&	2.21e-04	&	4.10e-03	&	8.56e-01	&	2.02e-03	&	7.51e-03	&	8.56e-01	&	6.49e-03	&	1.10e-02	&	8.87e-01\\
25000	&	8.0	&	5.24e-04	&	7.19e-03	&	7.29e-01	&	3.36e-03	&	1.47e-02	&	7.29e-01	&	1.46e-02	&	2.45e-02	&	8.72e-01\\
	&	12.0	&	8.07e-04	&	2.69e-02	&	6.08e-01	&	6.01e-03	&	4.29e-02	&	6.08e-01	&	3.37e-02	&	5.87e-02	&	8.31e-01\\
	&	16.0	&	2.39e-03	&	9.10e-02	&	5.81e-01	&	1.52e-02	&	1.33e-01	&	5.82e-01	&	9.88e-02	&	1.64e-01	&	7.55e-01\\
	&	24.0	&	2.91e-03	&	1.13e-01	&	6.43e-01	&	2.00e-02	&	1.63e-01	&	6.54e-01	&	1.29e-01	&	2.08e-01	&	7.61e-01\\
\hline
\enddata
\end{deluxetable*}

\begin{deluxetable*}{cccccccccc}
\tabletypesize{\footnotesize}
\tablecaption{Flux Data for Case A, i.e. without cooling\label{tab:fluxes_caseA}} 
\tablehead{
\colhead{Mass} & \colhead{Time} &  \colhead{Lst}  &  \multicolumn{6}{c}{Log(Flux/[mJy])} \\
 \colhead{ [$M_\odot$]} & \colhead{[yr]} & \colhead{[$L_\odot$]} & \colhead{500au} & \colhead{1000au} & \colhead{2000au} & \colhead{4000au} & \colhead{8000au} & \colhead{16000au} & \colhead{25000au \tablenotemark{a}}
 }
\startdata
\multicolumn{10}{c}{5.3 GHz} \\
1.4	&	4000	&	899.98	&	0.093$\pm$0.004	&	0.098$\pm$0.004	&	0.098$\pm$0.004	&	0.247$\pm$0.069	&	1.235$\pm$0.045	&	2.593$\pm$0.023	&	2.975$\pm$0.019	\\
2.0	&	9000	&	1317.53	&	0.217$\pm$0.003	&	0.249$\pm$0.003	&	0.250$\pm$0.003	&	0.250$\pm$0.003	&	0.250$\pm$0.003	&	2.117$\pm$0.071	&	2.649$\pm$0.034	\\
4.0	&	21000	&	1301.15	&	0.221$\pm$0.003	&	0.225$\pm$0.003	&	0.225$\pm$0.003	&	0.225$\pm$0.003	&	0.254$\pm$0.003	&	0.894$\pm$0.003	&	1.485$\pm$0.006	\\
8.0	&	39000	&	12499.0	&	0.397$\pm$0.019	&	0.411$\pm$0.019	&	0.419$\pm$0.020	&	0.422$\pm$0.020	&	0.481$\pm$0.015	&	1.655$\pm$0.045	&	2.655$\pm$0.032	\\
12.0	&	54000	&	44323.6	&	0.745$\pm$0.013	&	0.854$\pm$0.011	&	0.901$\pm$0.009	&	0.919$\pm$0.008	&	0.937$\pm$0.008	&	1.371$\pm$0.039	&	2.170$\pm$0.039	\\
16.0	&	68000	&	65461.5	&	0.556$\pm$0.008	&	0.702$\pm$0.006	&	0.787$\pm$0.005	&	0.830$\pm$0.004	&	0.861$\pm$0.004	&	0.897$\pm$0.004	&	1.713$\pm$0.066	\\
24.0	&	94000	&	84459.8	&	0.213$\pm$0.009	&	0.747$\pm$0.111	&	0.952$\pm$0.059	&	1.017$\pm$0.030	&	1.047$\pm$0.027	&	1.081$\pm$0.026	&	1.104$\pm$0.024	\\
\hline
\multicolumn{10}{c}{230.0 GHz} \\
1.4	&	4000	&	899.98	&	-0.063$\pm$0.004	&	-0.058$\pm$0.004	&	-0.058$\pm$0.004	&	0.097$\pm$0.074	&	1.121$\pm$0.054	&	2.538$\pm$0.022	&	2.874$\pm$0.020	\\
2.0	&	9000	&	1317.53	&	0.061$\pm$0.003	&	0.093$\pm$0.003	&	0.094$\pm$0.003	&	0.094$\pm$0.003	&	0.094$\pm$0.003	&	2.030$\pm$0.083	&	2.518$\pm$0.039	\\
4.0	&	21000	&	1301.15	&	0.074$\pm$0.003	&	0.078$\pm$0.003	&	0.078$\pm$0.003	&	0.078$\pm$0.003	&	0.106$\pm$0.003	&	0.735$\pm$0.003	&	1.322$\pm$0.006	\\
8.0	&	39000	&	12499.0	&	0.250$\pm$0.026	&	0.264$\pm$0.026	&	0.272$\pm$0.027	&	0.275$\pm$0.026	&	0.332$\pm$0.021	&	1.505$\pm$0.050	&	2.509$\pm$0.034	\\
12.0	&	54000	&	44323.6	&	1.358$\pm$0.455	&	1.390$\pm$0.429	&	1.404$\pm$0.418	&	1.410$\pm$0.413	&	1.416$\pm$0.409	&	1.602$\pm$0.292	&	2.128$\pm$0.122	\\
16.0	&	68000	&	65461.5	&	0.521$\pm$0.086	&	0.635$\pm$0.067	&	0.705$\pm$0.057	&	0.741$\pm$0.052	&	0.768$\pm$0.049	&	0.799$\pm$0.046	&	1.573$\pm$0.068	\\
24.0	&	94000	&	84459.8	&	0.573$\pm$0.337	&	0.875$\pm$0.242	&	1.003$\pm$0.201	&	1.048$\pm$0.179	&	1.069$\pm$0.173	&	1.093$\pm$0.166	&	1.109$\pm$0.161	\\
\enddata

\tablenotetext{a}{25000 au spans the entire simulation, which actually extends to $-15000\:\mathrm{au} <x<1500\:\mathrm{au}$, $0\:\mathrm{au}<z<25000\:\mathrm{au}$.}
\end{deluxetable*}

\setcounter{table}{0}
\setcounter{figure}{0}
\section{Effect of Varying Resolution}
\label{app:res}

\begin{figure*}
    \centering
    \includegraphics[width=1.0\textwidth]{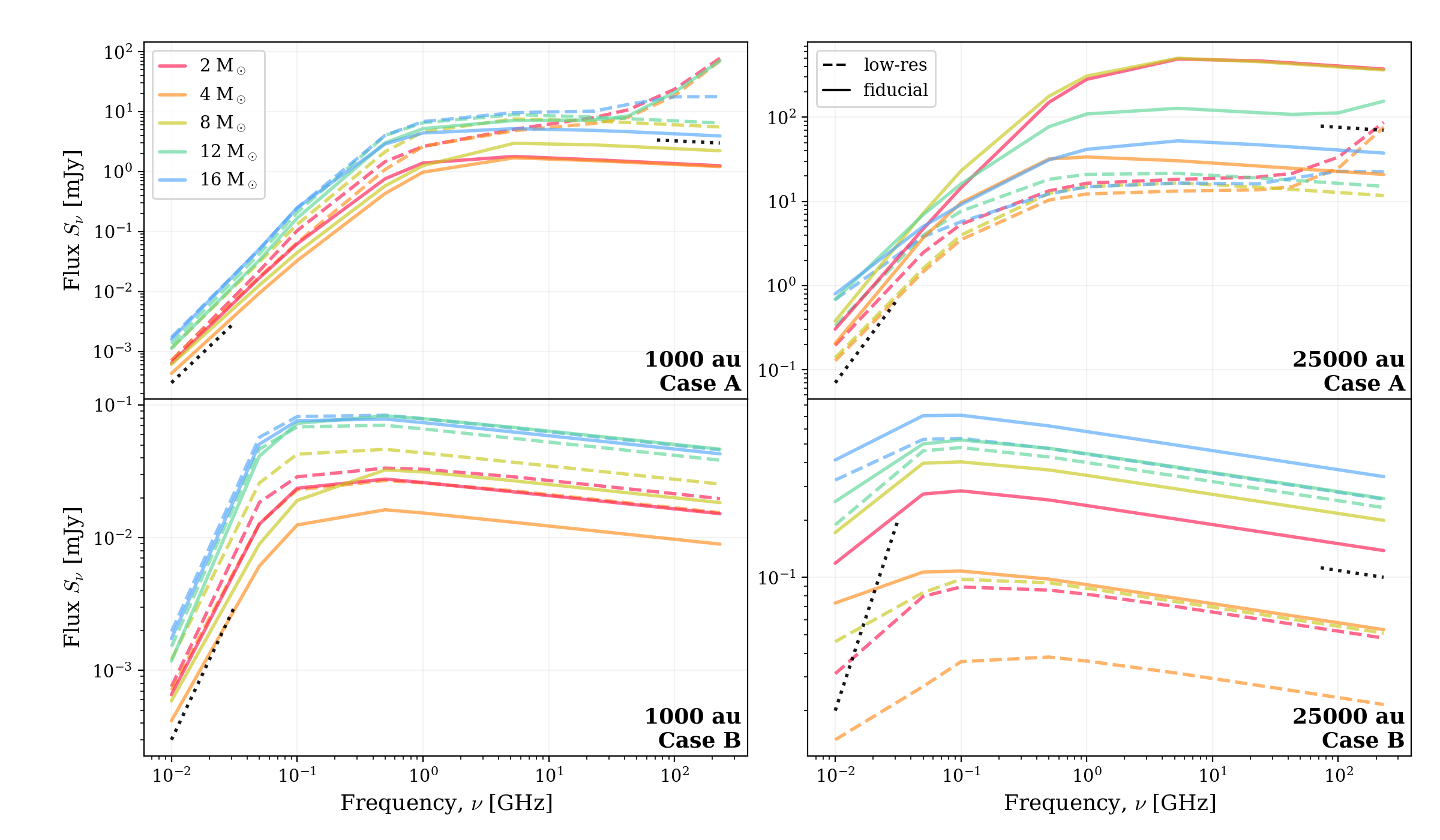}
    \caption{Free-free emission spectra for low (dashed lines) and fiducial (solid lines) resolutions, for Case A (without cooling) in the top row and Case B (with cooling) in the bottom row.
    The left column represents flux from the inner 1000 au$\times$1000 au and the right column represents flux from the entire domain.} 
    \label{fig:allres_spectra}
\end{figure*}

The fiducial models in this paper use simulations with a numerical resolution of $168\times280\times280$ cells logarithmically spaced over $25,000\ \mathrm{au}\times 30,000\ \mathrm{au} \times 30,000\ \mathrm{au}$. We conduct the same shock-modeling on a low resolution ($84\times140\times140$) 
version of this simulation, where all other parameters are kept the same, and present a comparison of the 
radio free-free emission spectra in Fig.~\ref{fig:allres_spectra}. 

These results indicate varying effects of resolution. For Case B in both the inner (1000~au) and outer (25,000~au) scale regions, there are relatively modest variations in flux between the low and fiducial resolutions, i.e., most fluxes are the same within a factor of 2, although some larger variations can arise, especially at lower frequencies when differing levels of optical depth can amplify differences. For Case A, larger differences can appear, some of which are also amplified by the effects of optical depth. Some differences in spectral slope at high frequencies indicate the presence of individual cells with local conditions that lead to being optically thick even at 100~GHz. Such cells tend to be less common in the fiducial resolution simulations.

\bibliography{references}{}
\bibliographystyle{aasjournal}



\end{document}